\documentclass[a4paper,fleqn]{cas-sc}

\usepackage[numbers]{natbib}

\usepackage{amsmath}
\usepackage{amssymb}
\usepackage{times}
\usepackage{braket}
\usepackage{blindtext}
\usepackage{color}

\begin{document}

\let\WriteBookmarks\relax
\def\floatpagepagefraction{1}
\def\textpagefraction{.001}

\renewcommand{\vec}[1]{\boldsymbol{#1}}
\newcommand{\change}[1]{\textcolor{black}{#1}}
\newcommand{\changeOT}[1]{\textcolor{black}{#1}}
\newcommand{\todo}[1]{\textbf{#1}}


\shorttitle{Beyond skyrmions: Review and perspectives of alternative magnetic quasiparticles}
\shortauthors{B{\"o}rge G{\"o}bel et~al.}

%

\title [mode = title]{Beyond skyrmions: Review and perspectives of alternative magnetic quasiparticles}    

\author[1,2]{B{\"o}rge G{\"o}bel}[orcid=0000-0003-4050-6869]
\ead{boerge.goebel@physik.uni-halle.de}
\cormark[1]

\author[1]{Ingrid Mertig}[orcid=0000-0001-8488-0997]

\author[3]{Oleg A. Tretiakov}[orcid=0000-0001-7283-6884]

\address[1]{Institut f\"ur Physik, Martin-Luther-Universit\"at Halle-Wittenberg, D-06099 Halle (Saale), Germany}
\address[2]{Max-Planck-Institut f\"ur Mikrostrukturphysik, D-06120 Halle (Saale), Germany}
\address[3]{School of Physics, The University of New South Wales, Sydney 2052, Australia}

\cortext[cor1]{Corresponding author}

\date{\today}

\begin{abstract}
Magnetic skyrmions have attracted enormous research interest since their discovery a decade ago. The non-trivial real-space topology of these nano-whirls leads to fundamentally interesting and technologically relevant consequences -- the skyrmion Hall effect of the texture and the topological Hall effect of the electrons. Furthermore, it grants skyrmions in a ferromagnetic surrounding great stability even at small sizes, making skyrmions aspirants to become the carriers of information in the future. Still, the utilization of skyrmions in spintronic devices has not been achieved yet, among other reasons, due to shortcomings in their current-driven motion. In this review, we present recent trends in the field of topological spin textures that go beyond skyrmions. The majority of these objects can be considered a combination of multiple subparticles, such as the bimeron, or the skyrmion analogues in different magnetic surroundings, such as antiferromagnetic skyrmions, as well as three-dimensional generalizations, such as hopfions. We classify the alternative magnetic quasiparticles -- some of them observed experimentally, others theoretical predictions -- and present the most relevant and auspicious advantages of this emerging field.
\end{abstract}



\maketitle

\tableofcontents


\section{Introduction} 

Over the last decades, information technology has become eminently relevant for our everyday lives. The recent conquest of modern IT applications, such as streaming services and cloud storages, has further intensified the demand for energy efficient data storage and manipulation. While current electronic solutions struggle to keep up with Moore's law~\cite{moore1965cramming}, new spintronic proposals have been suggested and may become relevant in the near future~\cite{waldrop2016chips}.

One of the most auspicious and anticipated data storage devices is the racetrack memory. Originally proposed for utilizing domain walls as the carriers of information~\cite{parkin2004shiftable,parkin2008magnetic,parkin2015memory}, the bits -- encoded by the presence or absence of the magnetic object -- are written, deleted, moved and read in a narrow track. This quasi-one-dimensional setup is stackable enabling the possibility for an innately three-dimensional data storage with drastically increased bit densities. This non-volatile concept without mechanically moving parts surpasses current random access memories (RAM) and hard disk drives (HDD) in terms of a lower energy consumption and faster access times~\cite{parkin2008magnetic}.

Besides spintronics, topological matter is an aspiring research field which is why this review is concerned with non-collinear spin textures. The most prominent example is the magnetic skyrmion~\cite{bogdanov1989thermodynamically}. This whirl-like nano-object was first observed a decade ago~\cite{muhlbauer2009skyrmion}. Its topological protection gives it an enormous stability even at small sizes, which makes it a potential carrier of information in future data storage devices, such as the racetrack nanodevices~\cite{sampaio2013nucleation,fert2013skyrmions,yu2017room}.

Besides great stability, the topological properties of skyrmions induce emergent electrodynamics, namely the topological Hall effect~\cite{bruno2004topological,neubauer2009topological,lee2009unusual} and the skyrmion Hall effect~\cite{zang2011dynamics,jiang2017direct,litzius2017skyrmion}. While the first -- an additional contribution to the Hall effect of electrons in the presence of a topologically non-trivial spin texture -- may become favorable for detecting skyrmions~\cite{hamamoto2016purely,maccariello2018electrical}, the skyrmion Hall effect leads to a  transverse deflection of skyrmions when they are driven by currents. This means that skyrmions are pushed towards the edge of the racetrack when a current is applied along the track, leading to pinning or even the loss of data. This is one of the reasons why no prototype of a skyrmion-based spintronic device exists today. 

While there will be ongoing research for improving the applicability of magnetic skyrmions in spintronic devices, several alternative nano-objects have been predicted and observed during the last 6 years. Some of them promise even greater advantages compared to conventional skyrmions, which is why research in this direction will be drastically \change{intensified} in the near future. 
In this review, we introduce and elaborate on these alternative magnetic quasiparticles. We establish a classification of the objects, explain methods to stabilize them, and compare their emergent electrodynamics to the case of conventional skyrmions. 

\begin{figure}[t!]
  \centering
  \includegraphics[width=\textwidth]{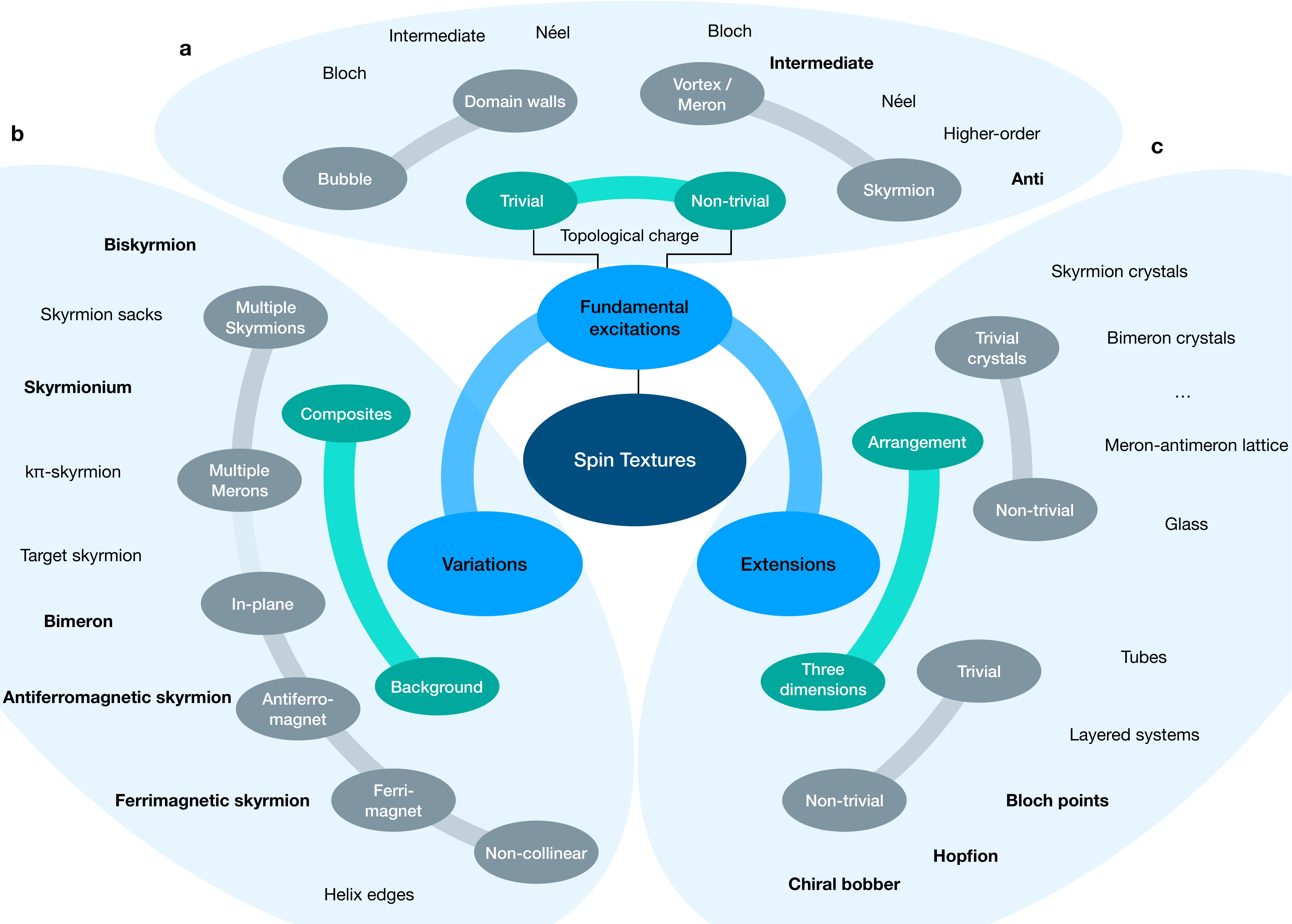}
  \caption{Classification of observed and predicted spin textures. (a) The fundamental excitations are the building blocks for all textures. Skyrmions and merons of different types have been observed as the topologically non-trivial excitations in magnets. (b) These object\change{s} can be combined or considered in a different magnetic background to form new, distinct quasiparticles. (c) Also, the fundamental and derived objects can be continued as periodic or non-periodic arrangements in two dimension\change{s} or can be continued along the third dimension trivially or non-trivially. The objects discussed more thoroughly in this review are typeset bold.}
  \label{fig:overview}
\end{figure}

\begin{figure}[t!]
  \centering
  \includegraphics[width=\textwidth]{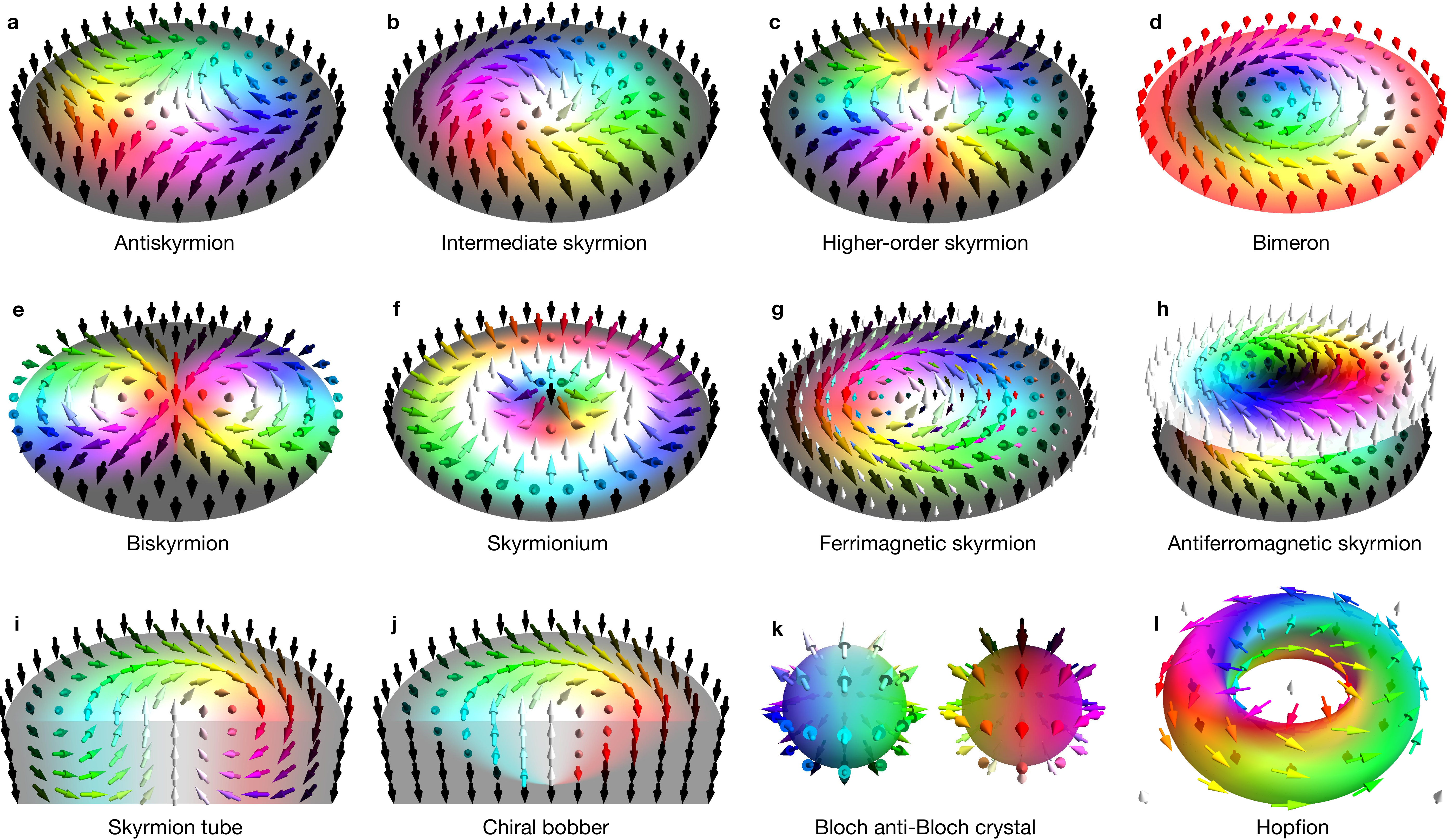}
  \caption{Overview of the discussed topologically non-trivial spin textures. The first \change{three} objects are different types of skyrmions, meaning skyrmions with various helicities and vortici\change{ti}es: (a) antiskyrmion with a vorticity $m=-1$ and a topological charge of $N_\mathrm{Sk}=-1$, (b) skyrmion with an intermediate helicity $\gamma=\pi/4$ between Bloch and N\'{e}el type skyrmions characterized by $N_\mathrm{Sk}=1$, (c) higher-order skyrmion with $N_\mathrm{Sk}=2$. (d) Shows a magnetic bimeron consisting of two merons. Alternatively, it can be understood as a skyrmionic excitation in an in-plane magnetized medium, here characterized by $N_\mathrm{Sk}=-1$. The middle row shows combinations of two skyrmions: (e) the biskyrmion with $N_\mathrm{Sk}=2$, (f) a skyrmionium with $N_\mathrm{Sk}=0$ and (g,h) ferrimagnetic and synthetic antiferromagnetic skyrmions for which the topological charges of the two subskyrmions compensate each other. The bottom row shows three-dimensional extensions of skyrmions: (i) skyrmion tubes (possibly with a varying helicity along the tube), (j) a chiral bobber as a discontinued skyrmion tube, (k) a pair of Bloch and anti-Bloch points constituting the building block of a three-dimensional crystal (hedgehog lattice), and (l) the hopfion. \change{The colored arrows represent the orientation of magnetic moments. White and black represent the positive and negative out-of-plane orientations, respectively. The different colors represent the varying in-plane orientations.}}
  \label{fig:particles}
\end{figure}

We begin by elaborating on conventional magnetic skyrmions (Sec.~\ref{sec:skyrmion}) to convey the differences to alternative magnetic quasiparticles later in the paper. We introduce how different types of skyrmions can be characterized topologically and geometrically (Sec.~\ref{sec:topology}), 
explain how they form lattices (Sec.~\ref{sec:lattice}), 
introduce various mechanisms for their stabilization (Sec.~\ref{sec:stability}), and mediate the emergent electrodynamics (Sec.~\ref{sec:emergent}). Thereafter, we characterize and discuss the alternative magnetic quasiparticles (Sec.~\ref{sec:alternative}). 
We distinguish three groups as visualized in Fig.~\ref{fig:overview}: the fundamental excitations in ferromagnets (including different types of skyrmions; Fig.~\ref{fig:overview}a), variations of these excitations (the combination of multiple excitations or excitations in other backgrounds than ferromagnets; Fig.~\ref{fig:overview}b), and extensions (for example periodic arrays of magnetic objects or innately three-dimensional spin textures; Fig.~\ref{fig:overview}c).
First, we address the objects that are closely related to skyrmions (Sec.~\ref{sec:related}), namely skyrmions with an arbitrary helicity~\cite{okubo2012multiple,garlow2019quantification} (Fig.~\ref{fig:particles}b) and antiskyrmions~\cite{nayak2017magnetic} (Fig.~\ref{fig:particles}a). Thereafter, we consider combinations of skyrmions (Sec.~\ref{sec:combination}): bimerons~\cite{kharkov2017bound,gobel2018magnetic,gao2019creation} (Fig.~\ref{fig:particles}d), biskyrmions~\cite{yu2014biskyrmion,gobel2019forming} (Fig.~\ref{fig:particles}e), skyrmioniums~\cite{zhang2016control,zhang2018real} (Fig.~\ref{fig:particles}f), antiferromagnetic skyrmions~\cite{barker2016static,zhang2016magnetic,legrand2019room} (Fig.~\ref{fig:particles}h) 
and ferrimagnetic skyrmions~\cite{woo2017current,kim2017self} (Fig.~\ref{fig:particles}g). 
Finally, we discuss 
three-dimensional objects, namely Bloch points~\cite{wild2017entropy,im2019dynamics} (Fig.~\ref{fig:particles}k),
chiral bobbers~\cite{rybakov2015new,zheng2018experimental} (Fig.~\ref{fig:particles}j) and hopfions~\cite{liu2018binding} (Fig.~\ref{fig:particles}l) in Sec.~\ref{sec:3d}. Other magnetic quasiparticles (shown in Fig.~\ref{fig:particles}) are briefly addressed and put in context while discussing the above objects. We conclude this review in Sec.~\ref{sec:conclusion}.

\section{Magnetic skyrmions}\label{sec:skyrmion}

Skyrmions have originally been predicted in the 1960s \change{by Tony Skyrme} in the context of particle physics. \change{He proposed a field-theoretical description of interacting pions and showed that particle-like solutions with attributes of baryons exist}~\cite{skyrme1961non,skyrme1962unified}. \change{Later it was shown that these solutions exhibit Fermi characteristics} while pions themselves are bosonic~\change{\cite{finkelstein1968connection,adkins1983static}}. The solitons, described by a non-linear sigma model, are the three-dimensional versions of what became known as skyrmions.

Today, skyrmions have been found in several fields of physics such as quantum Hall systems~\cite{sondhi1993skyrmions}, Bose-Einstein condensates~\cite{al2001skyrmions}, liquid crystals~\cite{fukuda2011quasi}, particle physics~\cite{adkins1983static}, string theory~\cite{vilenkin2000cosmic} and, as considered here, in magnetism~\cite{muhlbauer2009skyrmion}. 

In this context, a skyrmion can be considered as a two-dimensional object (Fig.~\ref{fig:topology}a), that is continued trivially along the third dimension (Fig.~\ref{fig:topology}c). Such skyrmion tubes or skyrmion strings have been observed for the first time in 2009 in MnSi by reciprocal-space measurements~\cite{muhlbauer2009skyrmion} and one year later using Lorentz transmission electron microscopy (LTEM)~\cite{yu2010real}. The magnetic textures of these objects were in agreement with what had been predicted twenty years earlier~\cite{bogdanov1989thermodynamically}: Magnetic skyrmions in a ferromagnetic medium are characterized by a continuously changing magnetization density which is oriented oppositely in its center compared to the surrounding leading to a non-trivial real-space topology.
These objects can occur as periodic lattices, like in the above publications, or as individual particles~\cite{bogdanov1994thermodynamically,yu2010real}.

The following discussion of conventional skyrmions is limited to their geometrical characterization, stabilizing mechanisms and emergent electrodynamics -- all of which are a prerequisite for understanding the physics of the alternative magnetic quasiparticles. For a discussion of conventional skyrmions that goes beyond these points, we refer to one of the many review articles~\cite{nagaosa2013topological,kang2016skyrmion,wiesendanger2016nanoscale, garst2017collective,finocchio2016magnetic,fert2017magnetic, jiang2017skyrmions,everschor2018perspective,zhou2019magnetic}.

\subsection{Topology and characterization} \label{sec:topology}

\begin{figure}[t!]
  \centering
  \includegraphics[width=0.65\columnwidth]{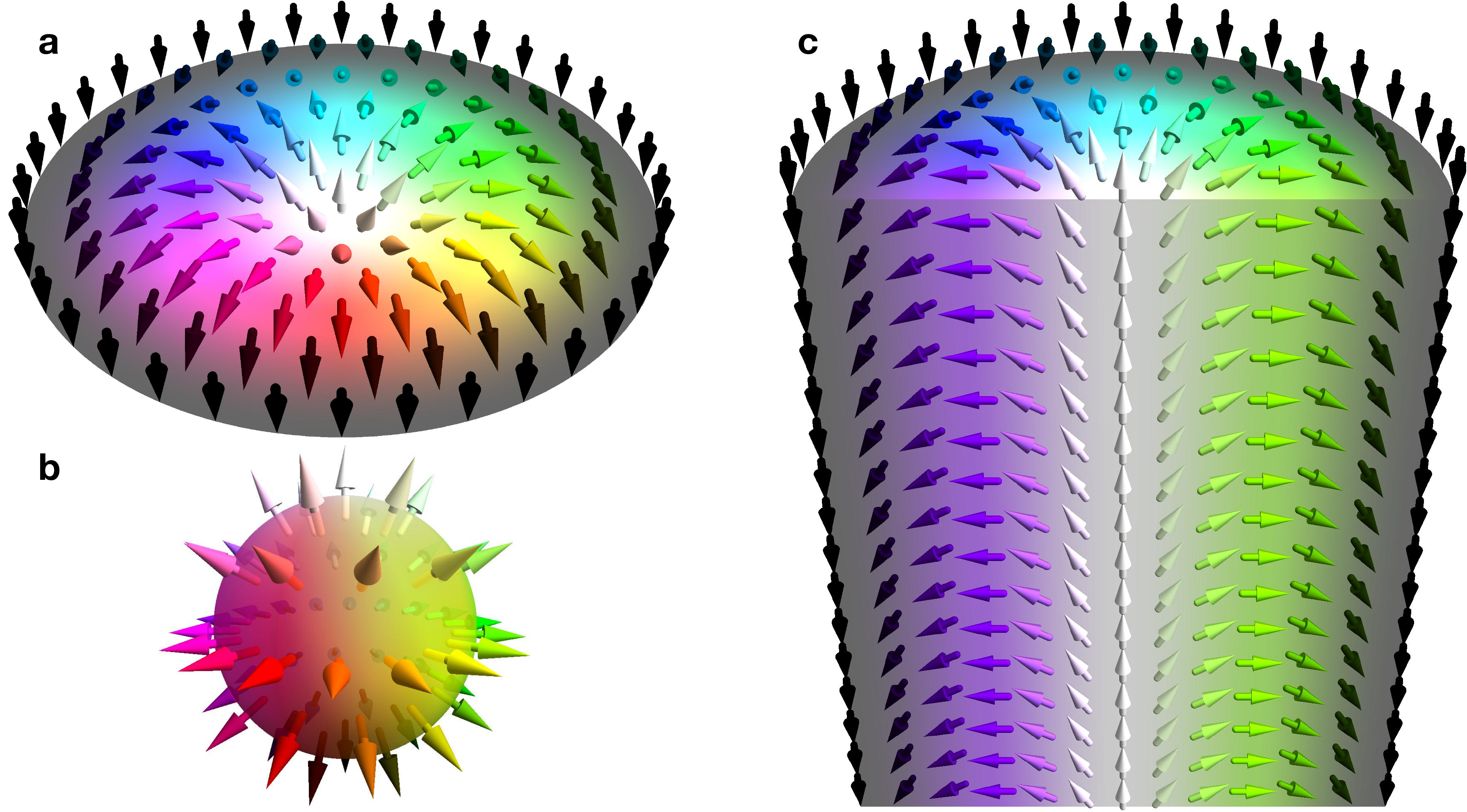}
  \caption{Magnetic skyrmions as fundamental topologically non-trivial excitations in ferromagnets. (a) A two-dimensional magnetic skyrmion as constructed from (b) a three-dimensional Bloch point or hedgehog by a stereographic projection. (c) In three dimensions, the skyrmion commonly extends trivially as a skyrmion tube.}
  \label{fig:topology}
\end{figure}

The topological character of a skyrmion can be comprehended by a stereographic projection: A two-dimensional skyrmion (Fig.~\ref{fig:topology}a) can be constructed by rearranging the magnetic moments of a three-dimensional hedgehog (also called Bloch point; see Fig.~\ref{fig:topology}b), where all moments on a sphere point along the radial direction. This sphere is opened at the bottom and flattened to a disk without changing the moments' orientations. The result is a topologically non-trivial magnetic object in two-dimensions. 

In a continuous picture, where a skyrmion consists of a magnetization density $\vec{m}(\vec{r})$, this skyrmion cannot be transformed to a ferromagnetic state without discontinuous changes in the density. This is a manifestation of the non-trivial real-space topology quantified by the topological charge 
\begin{equation}
N_\mathrm{Sk}=\int n_\mathrm{Sk}(\vec{r})\,\mathrm{d}^2r,
\end{equation}
which is as an integral over the topological charge density
\begin{equation}
n_\mathrm{Sk}(\vec{r})  = \frac{1}{4\pi} \vec{m}(\vec{r}) \cdot \left[ \frac{\partial \vec{m}(\vec{r})}{\partial x}  \times  \frac{\partial \vec{m}(\vec{r})}{\partial y}  \right]\label{eq:nsk}.
\end{equation}

The topological charge of a skyrmion can be determined more easily from its appearance due to the following transformation. One expresses the magnetization density in spherical coordinates with the azimuthal angle $\theta$ and the polar angle $\Phi$ and expresses the position vector in polar coordinates $\vec{r}=r(\cos\phi,\sin\phi)$. Exploiting the radial symmetry of the out-of-plane magnetization density $\theta=\theta(r)$, the topological charge reads~\cite{nagaosa2013topological}
\begin{equation}
\begin{split}
N_\mathrm{Sk}&=\frac{1}{4\pi}\int_0^\infty\mathrm{d}r\int_0^{2\pi}\mathrm{d}\phi\,\frac{\partial \Phi (\phi)}{\partial\phi}\frac{\partial\theta(r)}{\partial r}\sin\theta(r)\\
&=-\frac{1}{2}\cos\theta(r)\Big|_{r=0}^\infty\,\cdot\,\frac{1}{2\pi}\Phi(\phi)\Big|_{\phi=0}^{2\pi}.
\end{split}
\end{equation}
The out-of-plane magnetization of a skyrmion is reversed comparing its center with its confinement. This is quantified by the first factor, the polarity
\begin{equation}
p=-\frac{1}{2}\cos\theta(r)\Big|_{r=0}^\infty=\pm 1.
\end{equation}
The sign depends on the out-of-plane magnetization of the skyrmion host. Due to continuity of the magnetization density, the polar angle can only wrap around in multiples of $2\pi$ determining the second factor, the vorticity
\begin{equation}
m=\frac{1}{2\pi}\Phi(\phi)\Big|_{\phi=0}^{2\pi}=0,\pm 1,\pm 2, \ldots .
\end{equation}
The two-dimensional integral has been simplified to a product of the polarity and the vorticity~\cite{Tretiakov2007} 
\begin{equation}
N_\mathrm{Sk}=m\cdot p= \pm 1,\pm 2, \ldots,
\end{equation}
allowing only for integer topological charges for different types of skyrmions.

Besides their topological charges, different types of skyrmions can also differ by their in-plane magnetization. The polar angles of the position vector $\phi$ and the magnetization density $\Phi$ at each coordinate are related linearly,
\begin{equation}
\Phi=m\phi+\gamma,
\end{equation}
with an offset $\gamma$. This quantity is called helicity. 

With the polarity, vorticity and helicity we have established the three characterizing quantities for the different types of skyrmions, which can generally be expressed as
\begin{equation}
\vec{m}_\mathrm{sk}(\vec{r})=\begin{pmatrix} 
\left(\frac{x}{r}\cos\gamma-m\frac{y}{r}\sin\gamma\right) \sin\left(\frac{\pi}{r_0}r\right) \\
\left(\frac{x}{r}\sin\gamma + m\frac{y}{r}\cos\gamma\right)\sin\left(\frac{\pi}{r_0}r\right) \\
p\cos\left(\frac{\pi}{r_0}r\right)
\end{pmatrix}
\end{equation} 
\change{for $0<r<r_0$.} Note, that the out-of-plane magnetization profile is simplified as a cosine function (radius $r_0$) and that the exact profile depends on the interaction parameters, the sample geometry, defects, and the presence of other quasiparticles. 

As an example, the skyrmion in Fig.~\ref{fig:topology}a has a positive polarity $p=+1$ and vorticity $m=+1$ leading to a topological charge of $N_\mathrm{Sk}=+1$. Since the in-plane component of the magnetization is always pointing along the radial direction, the helicity in this case is $\gamma=0$. This type of skyrmion is called N\'{e}el skyrmion and is typically observed at interfaces~\cite{heinze2011spontaneous}. \change{Different to this type of skyrmions are} the skyrmions in MnSi (e.\,g. from the initial observation~\cite{muhlbauer2009skyrmion}). \change{They} are called Bloch skyrmions. There, the in-plane components of the magnetization density are oriented perpendicularly with respect to the position vector. This toroidal configuration is characterized by a helicity of $\gamma=\pm\pi/2$. 
In contrast to the polarity and the vorticity, the helicity is a continuous parameter allowing for skyrmions as intermediate states between Bloch and N\'{e}el skyrmions, as shown in Fig.~\ref{fig:particles}b.
Furthermore, the vorticity can in principle take any integer value constituting for example antiskyrmions for $m=-1$ (Fig.~\ref{fig:particles}a) or higher-order (anti)skyrmions for $|m|>1$ (Fig.~\ref{fig:particles}c). Out of this manifold, Bloch~\cite{muhlbauer2009skyrmion}, N\'{e}el skyrmions~\cite{heinze2011spontaneous} and skyrmions with an intermediate helicity~\cite{garlow2019quantification}, as well as antiskyrmions~\cite{nayak2017magnetic} have been observed experimentally. Higher-order skyrmions~\change{\cite{leonov2015multiply,ozawa2017zero,rozsa2017formation}} have been predicted.

\subsection{Skyrmion lattices}\label{sec:lattice}

\begin{figure}[t!]
  \centering
  \includegraphics[width=0.85\textwidth]{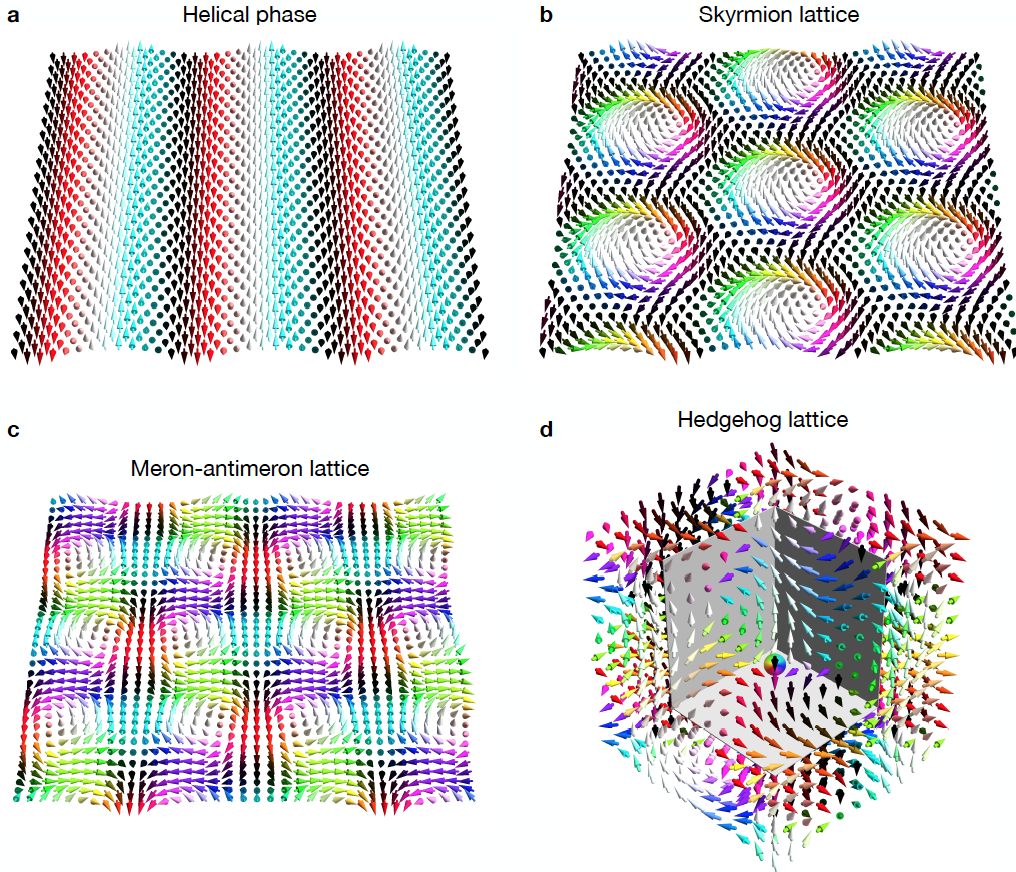}
  \caption{\change{Periodic spin textures.} (a) Helical phase, characterized by a single $Q$ vector. (b) Skyrmion lattice or skyrmion crystal constructed as a superposition of three helices with $Q$ vectors at a respective angle of $120^\circ$, taken from Ref. \cite{gobel2020emergent}. The skyrmions (vorticity $m=+1$) are of Bloch type (helicity $\gamma=-\pi/2$) and form a hexagonal superlattice, as for example in the B20 material MnSi \cite{yu2010real}. (c) Meron-antimeron lattice that consists of antimerons (vorticity $m=-1$) with a negative net magnetization (black) and merons (vorticity $m=+1$) with a positive net magnetization (white). The topological charge of both objects is $N_\mathrm{Sk}=+1/2$ giving the lattice a positive net topological charge. (d) Hedgehog lattice or Bloch anti-Bloch crystal formed by the superposition of three helices in three-dimensional space. In a continuous description, the magnetization density has singularities. These are Bloch or anti-Bloch points at which the magnetization is not defined. One of these points is highlighted by a colored sphere, similar to the Bloch point in Fig. \ref{fig:topology}b.}
  \label{fig:lattice}
\end{figure}

Skyrmions (and also alternative magnetic quasiparticles) tend to form lattices. In MnSi, for example, the ground state at zero temperature is a helical phase (Fig. \ref{fig:lattice}a), i.e. a spin spiral that has only one ordering vector $\vec{Q}$. However, at finite temperatures thermal fluctuations become important, so that the three spirals superimpose with each other at equivalent angles of $120^\circ$. Furthermore, the application of a magnetic field favors configurations with a net magnetization. The result is that a hexagonal superlattice of magnetic Bloch skyrmions (Fig. \ref{fig:lattice}b) becomes energetically more favorable than the helical phase for a small range of temperature and magnetic field \cite{muhlbauer2009skyrmion}. Mathematically, such a texture can be constructed as \cite{okubo2012multiple}
\begin{equation}
\begin{split}
\tilde{\vec{m}}_{xy}(\vec{r})&=\sum_{j=1}^3\sin(\vec{Q}_i\cdot \vec{r}+\theta_j)\,\vec{e}_j,\\
\tilde{m}_{z}(\vec{r})&=\sum_{j=1}^3\cos(\vec{Q}_i\cdot \vec{r}+\theta_j)\change{+m_0},
\end{split}
\end{equation}
where $\vec{e}_i$ are the unit vectors with $|\vec{e}_i|=1$, $\sum \vec{e}_i=0$, and with the phase factors $\cos(\sum\theta_i)=-1$. The magnetization profile is normalized $\vec{m}(\vec{r})=\tilde{\vec{m}}(\vec{r})/|\tilde{\vec{m}}(\vec{r})|$. \change{$m_0$ accounts for the fact that the stability of skyrmion crystals requires a finite magnetic field that tilts the moments out of the plane.} The three $\vec{Q}_i$ vectors are the ordering vectors of the three ground state spin spirals. Interestingly, it has been shown that in Co$_8$Zn$_9$Mn$_3$ the superposition of two $\vec{Q}_i$ vectors leads to the generation of a meron-antimeron lattice \cite{yu2018transformation} (Fig. \ref{fig:lattice}c). These are 'half (anti-) skyrmions' with a topological charge $N_\mathrm{Sk}=\pm 1/2$ as will be explained later. Moreover, if the $\vec{Q}_i$ vectors are non-coplanar, an innately three-dimensional hedgehog lattice forms like in MnGe \cite{kanazawa2011large,tanigaki2015real}. Recently, it has been found that these textures come in two varieties that can be characterized by either three $\vec{Q}_i$ vectors (pairwise perpendicular) or four $\vec{Q}_i$ vectors (spanning a tetrahedron) \cite{fujishiro2019topological}.

\subsection{Stabilizing mechanisms}\label{sec:stability}

Having discussed the different types of possible skyrmions and how lattices can form as the superposition of multiple helices mathematically, we will now address how these objects are physically stabilized in ferromagnets. In such materials the magnetic moments $\{\vec{s}_i\}$ are coupled via the exchange interaction
\begin{equation}
H_\mathrm{ex}=-\frac{1}{2}\sum_{ij}\change{J_{i,j}}\,\vec{s}_i\cdot\vec{s}_j.
\end{equation}
Typically, the nearest-neighbor interaction is dominant. In a ferromagnet it favors a parallel alignment ($J_{i,j}=J>0$ for $\braket{i,j}$ nearest neighbors; $J_{i,j}=0$ otherwise) of the magnetic moments. In a continuous approximation, this term is expressed as 
\begin{equation}
H_\mathrm{exchange}=\int\sum_{ij}\tilde{J}\left(\frac{\partial m_i}{\partial x_j}\right)^2\,\mathrm{d}^3r.
\end{equation}
In the continuous limit, a magnetic skyrmion would be stable due to the topological protection; it cannot be transformed into a uniform ferromagnet continuously, even though the ferromagnetic state would have the lower energy. However, since real skyrmions consist of magnetic moments and not a continuous density, this protection is not strict in nature~\cite{DeLucia2017} bringing forth the necessity of additional stabilizing interactions.

In theory, skyrmions have been stabilized by frustrated exchange interactions~\cite{okubo2012multiple}, four-spin interactions, dipole-dipole interactions~\change{\cite{garel1982phase}} and Dzyaloshinskii-Moriya interactions~\cite{nagaosa2013topological}. In the following, we will focus on the latter two cases, since they are by far the most relevant mechanisms in nature.

\subsubsection{Dzyaloshinskii-Moriya interaction}
The Dzyaloshinskii-Moriya interaction~\cite{dzyaloshinsky1958thermodynamic,moriya1960anisotropic} (DMI) 
\begin{equation}
H_\mathrm{DMI}=\frac{1}{2}\sum_{ij}\vec{D}_{ij}\cdot(\vec{s}_i\times\vec{s}_j),\label{eq:dmi}
\end{equation}
is a chiral interaction responsible for the stability of most experimentally observed skyrmions. It is an energy correction due to spin-orbital coupling under a broken inversion symmetry and can be considered an antisymmetric exchange interaction, $\vec{D}_{ij}=-\vec{D}_{ji}$. The $\vec{D}_{ij}$ are the Dzyaloshinskii-Moriya vectors, whose orientations account for the way the inversion symmetry is broken, satisfying the Moriya symmetry rules~\cite{moriya1960anisotropic}. \change{The full set of Lifshitz invariants and the corresponding stabilized skyrmions and antiskyrmions can be found in Ref~\cite{bogdanov2002magnetic}}. For example, at an interface of a magnet and a heavy metal, like Co/Pt, the DMI vectors are oriented parallel to the interfacial plane, perpendicular to the bond of the Co atoms~\cite{fert1980role}. In a continuous approximation this is expressed as~\cite{thiaville2012dynamics}
\begin{equation}
H_\mathrm{interface}=\int \tilde{D}\Big(m_x\frac{\partial m_z}{\partial x}-m_z\frac{\partial m_x}{\partial x}
 + m_y\frac{\partial m_z}{\partial y}-m_z\frac{\partial m_y}{\partial y}\Big)\mathrm{d}^3r.
\end{equation}
Such an interaction favors a canting of magnetic moments. As a consequence, N\'{e}el-type skyrmions may be stabilized, as first observed at the interface of Fe and Ir(111)~\cite{heinze2011spontaneous}. This type of DMI and the resulting skyrmions are well understood today. In multistack systems of magnetic and heavy metal materials, the effective DMI strength can be tuned~\cite{moreau2016additive} changing the size and the stability of N\'{e}el skyrmions~\cite{soumyanarayanan2017tunable}. Also, recently it has been demonstrated experimentally that the interfacial DMI can be enhanced by engineering a correlated roughness of the interfaces at the atomic length scales~\cite{samardak2020}.

Different samples lead to different DMI vectors, since the inversion symmetry is broken in a different way.
In Heusler materials, the layers are stacked in a way that the DMI vectors along the two bond directions have opposite signs 
\begin{equation}
H_\mathrm{aniso}=\int \tilde{D}\Big(m_x\frac{\partial m_z}{\partial x}-m_z\frac{\partial m_x}{\partial x}
-m_y\frac{\partial m_z}{\partial y}+m_z\frac{\partial m_y}{\partial y}\Big)\mathrm{d}^3r.\label{eq:DMIaniso}
\end{equation}
favoring antiskyrmions energetically~\cite{nayak2017magnetic}.
In B20 materials, such as MnSi, where the inversion symmetry is broken intrinsically, the DMI is expressed as
\begin{equation}
H_\mathrm{bulk}=\int \tilde{D}\Big(m_y\frac{\partial m_z}{\partial x}-m_z\frac{\partial m_y}{\partial x}+m_z\frac{\partial m_x}{\partial y}-m_x\frac{\partial m_z}{\partial y}+m_x\frac{\partial m_y}{\partial z}-m_y\frac{\partial m_x}{\partial z}\Big)\mathrm{d}^3r,
\end{equation}
leading to the stability of Bloch skyrmions~\cite{yu2010real}.

In principle, lower symmetric lattice structures or nano-structuring allow to generate further types of DMIs, leading to the stabilization of alternative topologically non-trivial excitations in magnets.

\subsubsection{Dipole-dipole interaction}

A second main mechanism for stabilizing skyrmions is the dipole-dipole interaction
\begin{equation}
H_{\mathrm{dd},ij}=-\frac{\mu_0}{4\pi}\left(3\frac{(\vec{s}_i\cdot\vec{r}_{ij})(\vec{s}_j\cdot\vec{r}_{ij})}{r_{ij}^5}-\frac{\vec{s}_i\cdot\vec{s}_j}{r_{ij}^3}\right).\label{eq:dipoledipole}
\end{equation}
While the DMI typically stabilizes skyrmions smaller than a few hundred nanometers, the dipole-dipole interaction can stabilize \change{larger} objects\change{; up to} several micrometer\change{s} in diameter~\cite{nagaosa2013topological}. Typically, these objects have an almost ferromagnetic center surrounded by a narrow domain wall. They are sometimes labeled `bubble' but are topologically equivalent to skyrmions, given that they have the same $N_\mathrm{Sk}$.
As a difference to the DMI, dipole-dipole interactions are achiral, meaning that they energetically favor Bloch skyrmions of both helicities $\gamma=\pm\pi/2$~\cite{malozemoff2016magnetic,gobel2019forming} allowing even for their coexistence~\cite{yu2012magnetic}.

\subsection{Emergent electrodynamics}\label{sec:emergent}

Magnetic skyrmions have successfully been generated and deleted in thin films (see reviews~\cite{nagaosa2013topological,kang2016skyrmion,wiesendanger2016nanoscale, garst2017collective,finocchio2016magnetic,fert2017magnetic, jiang2017skyrmions,everschor2018perspective} for different methods). In order to constitute an operating data storage device, they also have to be driven and read. In this section we will discuss the topological Hall effect and the current-driven motion of skyrmions. This will be the foundation for our discussion of alternative magnetic quasiparticles later in this review and will reveal why skyrmions may not be the optimal candidates for such spintronic devices. 

\begin{figure}[t!]
  \centering
  \includegraphics[width=0.65\columnwidth]{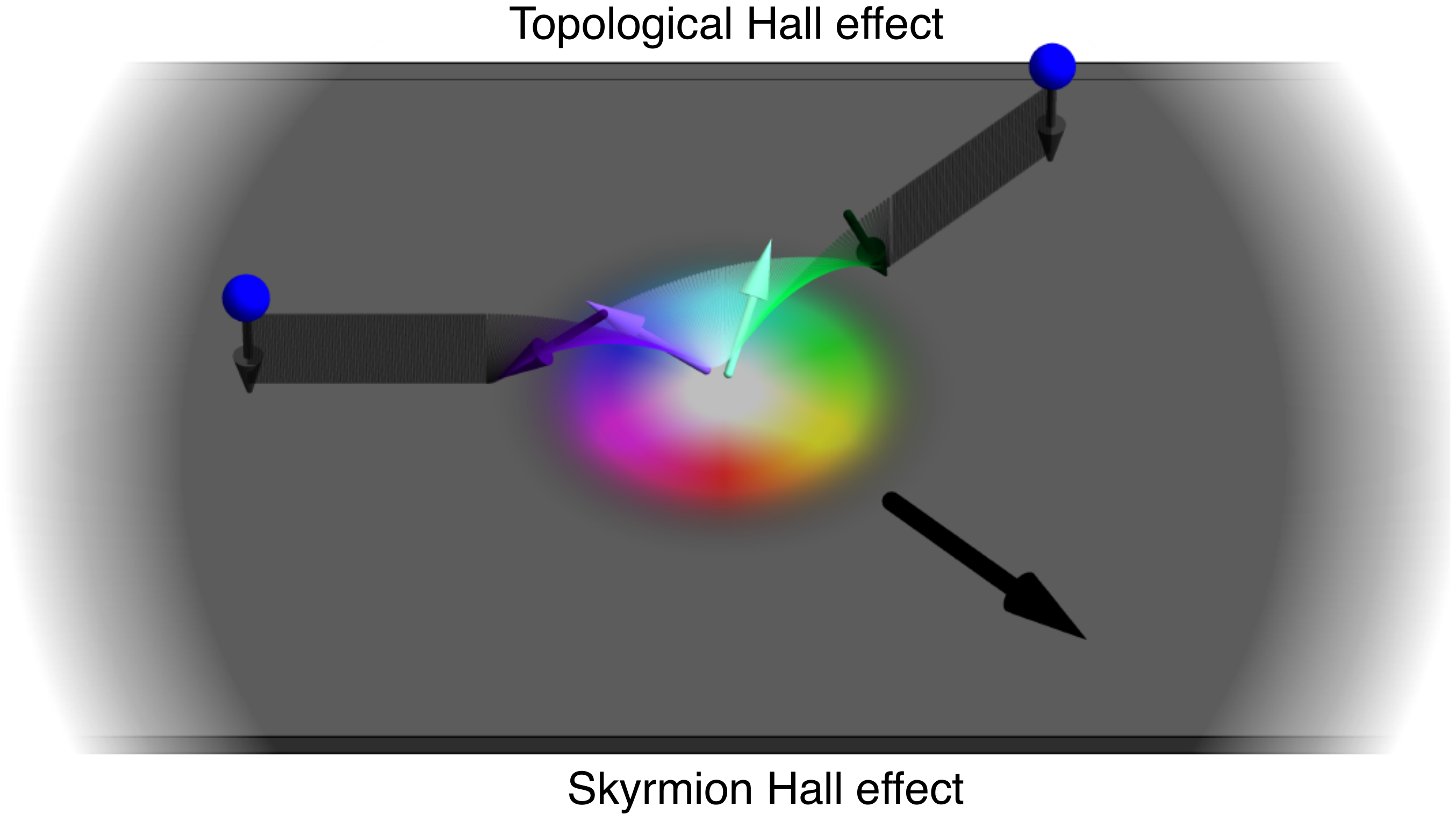}
  \caption{Emergent electrodynamics of a skyrmion. The topological Hall effect and the skyrmion Hall effect due to spin-transfer torque under application of an electric current are visualized. The current electrons \change{(blue)} and the skyrmion itself experience a transverse deflection, since the electron spin \change{(colored arrows)} aligns partially with the texture and generates a torque.}
  \label{fig:emergent}
\end{figure}

\subsubsection{Topological Hall effect}
The topological Hall effect of electrons is considered the hallmark of the skyrmion phase~\cite{bruno2004topological,neubauer2009topological,lee2009unusual,schulz2012emergent,li2013robust,hamamoto2015quantized,gobel2017THEskyrmion, gobel2017QHE,gobel2018family,hamamoto2016purely,maccariello2018electrical}. To measure it, an electric field $\vec{E}$ is applied to the skyrmion host, and the current density $\vec{j}$ is detected. According to Ohm's law, $\vec{E}=\rho\vec{j}$, the Hall effect of electrons is characterized by the transverse resistivity tensor element $\rho_{xy}$. For a skyrmion crystal, it is commonly considered as three superimposed contributions~\cite{neubauer2009topological}
\begin{equation}
\rho_{xy}=\rho_{xy}^\mathrm{HE}+\rho_{xy}^\mathrm{AHE}+\rho_{xy}^\mathrm{THE}\label{eq:hall}
\end{equation} 
that can be isolated due to their distinct proportionalities:
the ordinary Hall contribution~\cite{hall1879new} occurs due to an external magnetic field $\rho_{xy}^\mathrm{HE}\propto B_z$, the anomalous Hall contribution~\cite{nagaosa2010anomalous} is due to spin-orbit coupling and, usually, a net magnetization $\rho_{xy}^\mathrm{AHE}\propto M_z$, and the topological Hall contribution appears due to the presence of skyrmions or other topologically non-trivial spin textures $\rho_{xy}^\mathrm{THE}\propto \braket{n_\mathrm{Sk}}$.

The presence of a skyrmion leads to the emergence of an additional contribution to the Hall effect for the following reason: If an electron hops between two sites and reorients its spin, the original transfer integral $t$ accumulates a complex phase factor~\cite{nagaosa2013topological}. This factor is the analogue of the Peierls phase, which characterizes the magnetic field in the ordinary Hall effect. Therefore, in the case of topological spin textures, it can be related to an effective vector potential. In an adiabatic approximation, the corresponding field $\vec{B}_\mathrm{em}$ is called `emergent field'~\cite{nagaosa2013topological}
\begin{equation}
B_\mathrm{em,\alpha}(\vec{r})=\frac{1}{2}\epsilon_{\alpha\beta\gamma}\vec{m}(\vec{r})\cdot\left(\frac{\partial\vec{m(\vec{r})}}{\partial \beta}\times\frac{\partial\vec{m(\vec{r})}}{\partial \gamma}\right).
\end{equation}
In two dimensions it is proportional to the topological charge density
\begin{equation}
\vec{B}_\mathrm{em}(\vec{r})=\change{4}\pi n_\mathrm{Sk}(\vec{r})\,\vec{e}_z.
\end{equation}
This fictitious field can be used to easily relate the transverse deflection and the generation of a Hall effect  with the presence of skyrmions (Fig.~\ref{fig:emergent}).
As long as the electron spin and the texture are coupled strongly, the topological Hall effect is proportional to the number of skyrmions in the sample $\rho_{xy}^\mathrm{THE}\propto \braket{B_{\mathrm{em},z}}\propto N_\mathrm{Sk}$. 

In this sense, the topological Hall effect is very similar to the ordinary Hall effect. Consequently, since for small skyrmions the emergent field can be as large as thousands of Tesla, there are theoretical predictions of a quantized version of the topological Hall effect as well, very similar to the quantum Hall effect \cite{hamamoto2015quantized,gobel2017THEskyrmion}. Differences between the two effects stem from the fact that the emergent field is a virtual quantity that is not directly measurable. In particular, it has an opposite orientation for spin-up and spin-down electrons, different from an actual magnetic field. Consequently, when the ordinary Hall effect changes its sign upon reoccupation of electron and hole states in a system (e.\,g. by changing the temperature or by applying a gate voltage), the topological Hall effect does not necessarily have to follow this behavior when spin-up and spin-down states are present \cite{raju2020formation}.

For skyrmions, as presented above, the three important effects -- ordinary Hall effect, anomalous Hall effect and topological Hall effect -- are caused by vectorial quantities that are all oriented out of the plane \change{($z$)}, along the skyrmion tube -- the external field \change{$\vec{B}$}, the net magnetization \change{$\vec{M}$}, and the emergent field \change{$\vec{B}_\mathrm{em}$}. However, these three quantities are not necessarily parallel to each other for alternative topological magnetic nano-objects\change{: e.\,g., we will show in Sec. \ref{sec:combination} that bimerons have an in-plane magnetization but an out-of-plane emergent field}. This allows to decouple the different contributions to the Hall effect of electrons \change{in a three-dimensional bulk sample}, because \change{the effects} are expected to appear in different resistivity tensor elements \change{$\rho_{ij}$; the $ij$ plane is perpendicular to the respective vector  $\vec{B}$, $\vec{M}$ or $\vec{B}_\mathrm{em}$}. This is one of the main motivations to investigate alternative magnetic quasiparticles. Later in this paper we will present the magnetic bimeron and the magnetic hopfion as two of such examples (shown in Fig. \ref{fig:differentemergent}).

\change{It has to be noted that it is hard to confirm experimentally, whether or not the topological Hall effect can be attributed to the emergent field of chiral spin textures. While in many samples the measured Hall signal is in good agreement with the number of detected skyrmions observed \cite{maccariello2018electrical}, this is not necessarily attributed to the skyrmions' emergent field. For example, the drop in the average magnetization due to the presence of skyrmions is also proportional to the number of skyrmions. For this reason, also in a theoretical description, it may not be sufficient to only employ the emergent field approach and more elaborate tight-binding or ab-initio approaches should be utilized. This is, however, difficult computationally due to the large number of magnetic moments in a skyrmion.} 

\change{Furthermore, b}esides the ordinary, anomalous and topological Hall effects, it is worth mentioning that several other origins of Hall effects have been identified over the recent years. \change{Most relevant for these systems seems to be} the chiral Hall effect \cite{lux2018engineering,lux2020chiral} caused by the gradient of the magnetization density (instead of its square as in the topological Hall effect). \change{Also} the anomalous Hall effect in several coplanar Kagome magnets without net magnetization \cite{chen2014anomalous,kubler2014non,busch2020microscopic, nakatsuji2015large,nayak2016large} and the crystal Hall effect for which a set of time reversal and spatial symmetries is broken by the arrangement of non-magnetic atoms in the crystal \cite{vsmejkal2019crystal,feng2020observation} \change{have been predicted and measured in magnetic systems}. Since their influence on conventional skyrmions is not yet well explored we abstain from making predictions on how these contributions affect alternative magnetic quasiparticles. However, for skyrmions, the experiments are well explained without these contributions. Still, these contributions could have a considerable quantitative influence.

\subsubsection{Skyrmion Hall effect}

The non-trivial real-space topology of skyrmions also becomes apparent in the current-driven motion of the skyrmions themselves. One typically discusses two scenarios: the motion under spin-transfer torque (STT)~\cite{zhang2004roles}, where a spin-polarized current goes through the spin texture and reorients the magnetic moments of a skyrmion (Fig.~\ref{fig:emergent}), and the motion under spin-orbit torque (SOT)~\cite{slonczewski1996current,zhang2002mechanisms,shpiro2003self, Ado2017}, where a spin accumulation created by an electric current in the presence of spin-orbit interaction exerts a torque $\vec{\tau}$ on the skyrmion texture. 
\change{The SOT scenario is typically characterized by the out-of-plane (field-like) torque $\propto \vec{m}\times\vec{s}$ and the in-plane (anti-damping-like) torque $\propto (\vec{m}\times\vec{s})\times\vec{m}$, that will be explained in the following. In the STT scenario, the torque is determined by the gradient of the magnetization density. One considers the adiabatic torque $\propto (\vec{j}\cdot\nabla)\vec{m}$ and the non-adiabatic torque $\propto \vec{m}\times(\vec{j}\cdot\nabla)\vec{m}$. These torques} reorient the magnetic moments. This collective reorientation can be identified with a motion of the skyrmions. 
The SOT mechanism turns out to be more efficient~\cite{sampaio2013nucleation}, since electron spins and magnetic moments can have a large misalignment leading to larger torques. Still, it has been predicted~\cite{sampaio2013nucleation} and observed~\cite{jiang2017direct,litzius2017skyrmion} that the skyrmions do not move parallel to the current, but experience a transverse deflection in this case. This phenomenon is called the skyrmion Hall effect and originates in the topological charge of the skyrmions.

For the given reasons, in this review we focus on the SOT setup and will only occasionally mention the spin-transfer torque. Typically, one considers a bilayer of a ferromagnet, potentially hosting skyrmions, and a heavy metal. The applied current mainly flows in the heavy metal, where the spin Hall effect generates a pure spin current along the perpendicular direction with spins oriented perpendicular to both currents. \change{In a racetrack geometry, the charge current flows along the racetrack, the spin current is oriented along the out-of-plane direction and the injected spins are perpendicular to both directions.} 

The reorientation of the magnetic moments can be modeled by the Landau-Lifshitz-Gilbert (LLG) equation~\cite{landau1935theory,gilbert1955lagrangian, slonczewski1996current}
\begin{equation}
\frac{\partial \vec{m}}{\partial t}=-\gamma\vec{m}\times\vec{B}_\mathrm{eff}+\alpha\vec{m}\times\frac{\partial\vec{m}}{\partial t}+\tau.\label{eq:llg}
\end{equation}
It is written here in the micromagnetic formulation, where the magnetization density has been discretized in small volumes with a normalized magnetization $\vec{m}_i$ (the index has been dropped for simplicity). The first term on the right side describes the precession of each normalized magnetic moment around its space- and time-dependent effective magnetic field 
\begin{equation}
\vec{B}_\mathrm{eff}=-\frac{1}{M_s}\frac{\delta F}{\delta \vec{m}},
\end{equation}
characterized by the free energy functional $F$ accounting for all magnetic interactions. $\gamma_e=\gamma/\mu_0=1.760\times 10^{11}\,\mathrm{T}^{-1}\mathrm{s}^{-1}$ quantifies the gyromagnetic ratio of an electron.
The second term is the Gilbert damping quantified by the dimensionless parameter $\alpha$, leading to an alignment of the magnetic moment with its effective field  after a certain propagation time without external perturbations.
The last term is the torque $\tau$. In the spin-orbit torque scenario it is given by the out-of-plane and in-plane torques. The out-of-plane torque term may change the shape of a skyrmion \change{thereby affecting the motion of the skyrmion}, but does not add \change{an individual term to the} effective equation of motion \change{[Eq. \eqref{eq:thielesot} presented in the following]}, which is why we consider only the in-plane torque term in the following. It reads~\cite{khvalkovskiy2013matching}
\begin{equation}
\tau=\frac{\gamma_e \hbar}{2ed_zM_s}\theta_\mathrm{SH}j[(\vec{m}\times\vec{s})\times\vec{m}],
\end{equation}
where $d_z$ is the thickness of the magnetic layer, $M_s$ is the saturation magnetization and $\theta_\mathrm{SH}$ is the spin-Hall angle describing the spin current $\theta_\mathrm{SH}j$ with spin orientation $\vec{s}$.

Numerically solving this equation for a skyrmion in a ferromagnet in the presence of an applied current leads to a motion of the skyrmion partially along the current direction (along the track), but also partially towards the track's edge. This transverse component is due to the topological charge of the skyrmion as can be found by considering the Thiele equation~\cite{thiele1973steady,Tretiakov2008,Clarke2008,sampaio2013nucleation,gobel2018overcoming}
\begin{equation}
b\,\vec{G}\times\vec{v}-b\underline{D}\alpha\vec{v}-Bj\underline{I}\vec{s}=\nabla U(y)\label{eq:thielesot}.
\end{equation}
This effective equation of motion can be derived from the LLG equation by assuming a perfectly rigid shape of the magnetic object and by considering only its center coordinate (with velocity $\vec{v}$). 
The non-collinearity of the object is condensed in the gyroscopic vector $\vec{G}$, the dissipative tensor $\underline{D}$ and the torque tensor $\underline{I}$, which are calculated as follows in two-dimensions (for a complete derivation see \cite{gobel2020emergent})
\begin{equation}
\begin{split}
\vec{G}&=-4\pi N_\mathrm{Sk}\vec{e}_z\\
D_{ij}&=\int\partial_{i}\vec{m}(\vec{r})\cdot\partial_{j}\vec{m}(\vec{r})\,\mathrm{d}^2r,\\
I_{ij}&=\int[\partial_{i}\vec{m}(\vec{r})\times\vec{m}(\vec{r})]_{j}\,\mathrm{d}^2r.
\end{split}
\end{equation}
The non-collinear object is then condensed to a single point.

A circular N\'{e}el skyrmion for example is characterized by a topological charge of $N_\mathrm{Sk}=p=\pm 1$, a diagonal dissipative tensor with only $D_{xx}=D_{yy}\neq 0$ and an antisymmetric torque tensor with only $I_{xy}=-I_{yx}\neq 0$. For injected spins $\vec{s}\parallel \pm\vec{y}$ this yields a skyrmion Hall angle of 
\begin{equation}
\tan\theta_\mathrm{Sk}=\tan\frac{v_y}{v_x}=\frac{-4\pi N_\mathrm{Sk}}{\alpha D_{xx}}
\end{equation}
in the absence of the potential gradient $\nabla U$. In this case, it makes sense to think that the deflection of a skyrmion is caused by its topological charge $N_\mathrm{Sk}$. \change{Also, we would like to mention that the out-of-plane torque term, that we neglected in the LLG equation \eqref{eq:llg}, would not lead to a new term in the Thiele equation \eqref{eq:thielesot} but it can influence the magnetization profile of a skyrmion. For example, it can lead to anisotropic deformations that break the relations $D_{xx}=D_{yy}$ and $I_{xy}=-I_{yx}$ and may therefore lead to considerable effects on the skyrmion propagation.}

Now that we have established the general equation of motion valid for two-dimensional magnetic quasiparticles, we can systematically think of ways to suppress the skyrmion Hall effect. We begin our discussion from the right side of the Thiele equation \eqref{eq:thielesot}. 
The term $\nabla U$ characterizes the force resulting from the interaction potential of a nano-object with the confinement of the sample, defects or other nano-objects. There are several approaches to utilize this interaction, such as adding a high-anisotropy material at the racetrack's confinement \cite{lai2017improved} or preparing a path using defects with a repulsive skyrmion-defect interaction \cite{fernandes2018universality}. These approaches mitigate the consequences of the skyrmion Hall effect, but cannot suppress its origin which is why their applicability is limited to low current densities. For N\'{e}el skyrmions the only other possibility, suggested by the Thiele equation, is to tune the injected spin orientation $\vec{s}$.
This can, for example, be done by attaching a thin magnetic layer on top of the skyrmion host \cite{zhang2015skyrmion}. If a current passes from the perpendicular direction, it becomes spin polarized with $\vec{s}$ parallel to the magnetization in that layer. In principle, this allows to tune $\vec{s}$ so that the driving force $-Bj\underline{I}\vec{s}$ is rotated and its transverse component compensates the gyroscopic force $-4\pi N_\mathrm{Sk}b\,\vec{e}_z\times\vec{v}$.  However, it is quite difficult to uniformly control the magnetization of such a layer without affecting the magnetization of the skyrmion hosting layer. 
More promising could be the control of $\vec{s}$ using the SOT scenario with low-symmetric heavy metals \cite{gobel2018overcoming}: It has been predicted that in triclinic, monoclinic and trigonal crystal systems, as well as in certain tetragonal or hexagonal crystal systems like \change{Sc, Hf, Ti \cite{freimuth2010anisotropic},} Pt$_3$Ge, Au$_4$Sc or (Au$_{1-x}$Pt$_x$)Sc \cite{wimmer2015spin}, the spin Hall effect generates a spin current with spins oriented partially along the current direction and not pairwise perpendicular to the spin and charge currents \cite{seemann2015symmetry}. The manipulated $\vec{s}$ has the same consequences as discussed above and micromagnetic simulations have shown that N\'{e}el skyrmions can be driven without a skyrmion Hall effect at velocities of several hundred meters per second \cite{gobel2018overcoming}. 

All of the above presented approaches to mitigate or suppress the skyrmion Hall effect are interesting theoretical predictions that are, however, difficult to realize. It may be more promising to utilize nano-objects that have an innately suppressed skyrmion Hall effect. This is possible for objects with $N_\mathrm{Sk}=0$ or for objects with modified tensors $\underline{I}$ and $\underline{D}$ compared to N\'{e}el skyrmions, for example due to a broken rotational symmetry. Furthermore, for three-dimensional objects the gyroscopic vector $\vec{G}$ is not necessarily along $\vec{e}_z$ anymore. This is one main motivation for considering alternative magnetic quasiparticles.

\subsubsection{Related effects}

Before we start our detailed discussion of the alternative magnetic nano-objects, we want to briefly mention a few other effects that are caused by the topological charge of non-collinear spin textures. These effects are directly related to the above presented topological Hall effect of electrons or the current-driven motion of spin textures.

As explained, the finite topological charge of skyrmions gives rise to a topological Hall effect of electrons. Directly related to this effect is the orbital magnetization. It is an additional contribution to the magnetization that arises due to the circulation of conduction electrons in a non-collinear spin texture or in the presence of spin-orbit coupling \cite{xiao2005berry}. Consequently, this effect has also been predicted for skyrmions \cite{dos2016chirality,lux2018engineering, gobel2018magnetoelectric}. In an insulator, the effect is directly tied to the Hall conductivity and determines its slope in dependence of the energy
\begin{equation}
\frac{\partial}{\partial E}M_z^\mathrm{orb}(E)=\frac{1}{2e}[\sigma_{yx}(E)-\sigma_{xy}(E)].
\end{equation}
It is expected that other topologically non-trivial nano-objects exhibit this effect as well.

The second effect that we would like to mention is the topological spin Hall effect. As was explained above, the electrons that are deflected transversally in the topological Hall effect are (partially) spin polarized, since they align with the non-collinear texture. This means that not only charge is transported, but spin is as well \cite{yin2015topological,gobel2018family}. As we show later in this review, there exist even nano-objects in an antiferromagnetic background that exhibit a pure topological spin Hall effect since the charge transport is compensated \cite{buhl2017topological,gobel2017afmskx,akosa2018theory}.

Finally, we would like to mention that skyrmions and other nano-objects may also interact not only with electrons but also with other (quasi-)particles. For example, skyrmions can be driven by magnons \cite{lin2014ac,mochizuki2014thermally}. Also, there exist topological versions of the magnon Hall effect \cite{schutte2014magnon,mook2017magnon} with magnonic edge states arising from skyrmion crystals \cite{roldan2016topological,diaz2020chiral}. However, in the following we restrict our review mainly to the topological Hall effect of electrons and the current-driven motion as these are the most intensively discussed emergent electrodynamic effects.


\section{Alternative magnetic quasiparticles} \label{sec:alternative}

In this main section of the review, alternative magnetic quasiparticles are introduced and discussed concerning their perspective for spintronic applications, mainly for racetrack memories. In this regard, we will refer to the fundamentals established in the last section. We characterize the different types of objects related to skyrmions, address their stability and their emergent electrodynamics.

\paragraph*{Fundamental excitations in ferromagnets.} 
As presented in Fig.~\ref{fig:overview}, we distinguish the fundamental excitations in ferromagnets from their variations and extensions. In panel a of Fig.~\ref{fig:overview} the fundamental excitations are shown. They can be separated into topologically trivial and non-trivial objects. The latter comprise merons and skyrmions of different varieties. As presented in the last section, topological excitations with a topological charge of $\pm 1$ have been found as N\'{e}el skyrmion~\cite{heinze2011spontaneous}, Bloch skyrmion~\cite{muhlbauer2009skyrmion}, intermediate skyrmions~\cite{garlow2019quantification} and antiskyrmion~\cite{nayak2017magnetic} (Fig.~\ref{fig:particles}a). Furthermore, higher-order skyrmions~\cite{ozawa2017zero} (Fig.~\ref{fig:particles}c) have been predicted. Most attractive application-wise are the intermediate skyrmions and antiskyrmion, since they can be moved without the occurrence of a skyrmion Hall effect~\cite{kim2018asymmetric,huang2017stabilization}, as will be presented in Sec.~\ref{sec:related}.

\paragraph*{Variations of the topological excitations.}
In Sec.~\ref{sec:combination} we discuss the combinations of skyrmions or merons to form new particles. Very promising are the combinations of two skyrmions with opposite topological charges: the antiferromagnetic skyrmion~\cite{barker2016static,legrand2019room} and the skyrmionium~\cite{zhang2016control,zhang2018real}, which are predicted to move without a skyrmion Hall effect. 
 
As shown in Fig.~\ref{fig:overview}b, some of the composite objects can ambivalently be considered as skyrmionic excitation in different magnetic backgrounds. The bimeron~\cite{kharkov2017bound,gobel2018magnetic,gao2019creation} (Fig.~\ref{fig:particles}d), for example, is on one side the combination of a meron and an antimeron, but also is a skyrmion in a ferromagnet which is magnetized in-plane. Likewise, the antiferromagnetic skyrmion~\cite{barker2016static,legrand2019room} (Fig.~\ref{fig:particles}h) is the combination of two ferromagnetic skyrmions with mutually reversed spins, but also is the fundamental topological excitation in a collinear antiferromagnet. The same holds for ferrimagnetic skyrmions~\cite{woo2017current,kim2017self} (Fig.~\ref{fig:particles}g) in ferrimagnets.

\paragraph*{Extensions of the topological excitations.}
The fundamental objects as well as their variations can be extended in the sense that they can be arranged in a (pseudo-) two-dimensional system, or that they can be extended along the third spatial dimension, see Fig.~\ref{fig:overview}c. 

The two-dimensional arrangements are typically rather trivial: Periodic crystals of skyrmions, antiskyrmions, bimerons, biskyrmions, antiferromagnetic skyrmions and other particles can form. However, such an arrangement can also be highly non-periodic like in a skyrmion glass~\cite{yu2015variation,hoshino2018theory}, or the arrangement can consist of multiple objects like for the meron-antimeron lattice, as observed recently~\cite{yu2018transformation}. 

However, more relevant for this review seem to be the three-dimensional extensions that are discussed in Sec.~\ref{sec:3d}. Skyrmion tubes (Fig.~\ref{fig:particles}i), chiral bobbers~\cite{rybakov2015new,zheng2018experimental} (Fig.~\ref{fig:particles}j), Bloch anti-Bloch crystals~\cite{kanazawa2011large} (Fig.~\ref{fig:particles}k), and hopfions~\cite{liu2018binding} (Fig.~\ref{fig:particles}l) are the most prominent candidates.


\subsection{Different types of skyrmions} \label{sec:related}
Skyrmions are characterized by the polarity $p$, vorticity $m$ and helicity $\gamma$. All types of skyrmions experience a skyrmion Hall effect under STT but behave differently under SOT. For example, the missing rotational symmetry of antiskyrmions brings about an anisotropic skyrmion Hall effect, which is highly relevant for racetrack applications. 

The vorticity can also have integer values larger than 1. This characterizes higher-order skyrmions and antiskyrmions with $|N_\mathrm{Sk}|>1$. These objects exhibit a topological Hall effect and a skyrmion Hall effect in the STT scenario just like skyrmions~\cite{xia2019current,weissenhofer2019orientation} but their rotational symmetry is broken, similar to antiskyrmions.

Furthermore, in this class of magnetic quasiparticles the fundamental excitations in in-plane magnetized samples are worth to be mentioned: merons (or vortices) which are closely related to skyrmions~\cite{kosevich1990magnetic}. These objects are configurated like skyrmions near their centers but the magnetic moments at the edges of the particles do not point into the opposite out-of-plane direction compared to the centers but along in-plane directions, which is often related to the magnetic anisotropy. The azimuthal angle changes only by $\pi/2$ giving these objects a polarity and a topological charge of $\pm 1/2$. These objects become relevant for forming non-trivial textures like the bimeron~\cite{kharkov2017bound,gobel2018magnetic,gao2019creation} or the meron-antimeron crystal~\cite{yu2018transformation} but are less relevant themselves, since they are always situated in a coplanar but non-collinear in-plane magnet and cannot be created as individual objects in a ferromagnet.

\begin{figure}[t!]
  \centering
  \includegraphics[width=0.88\textwidth]{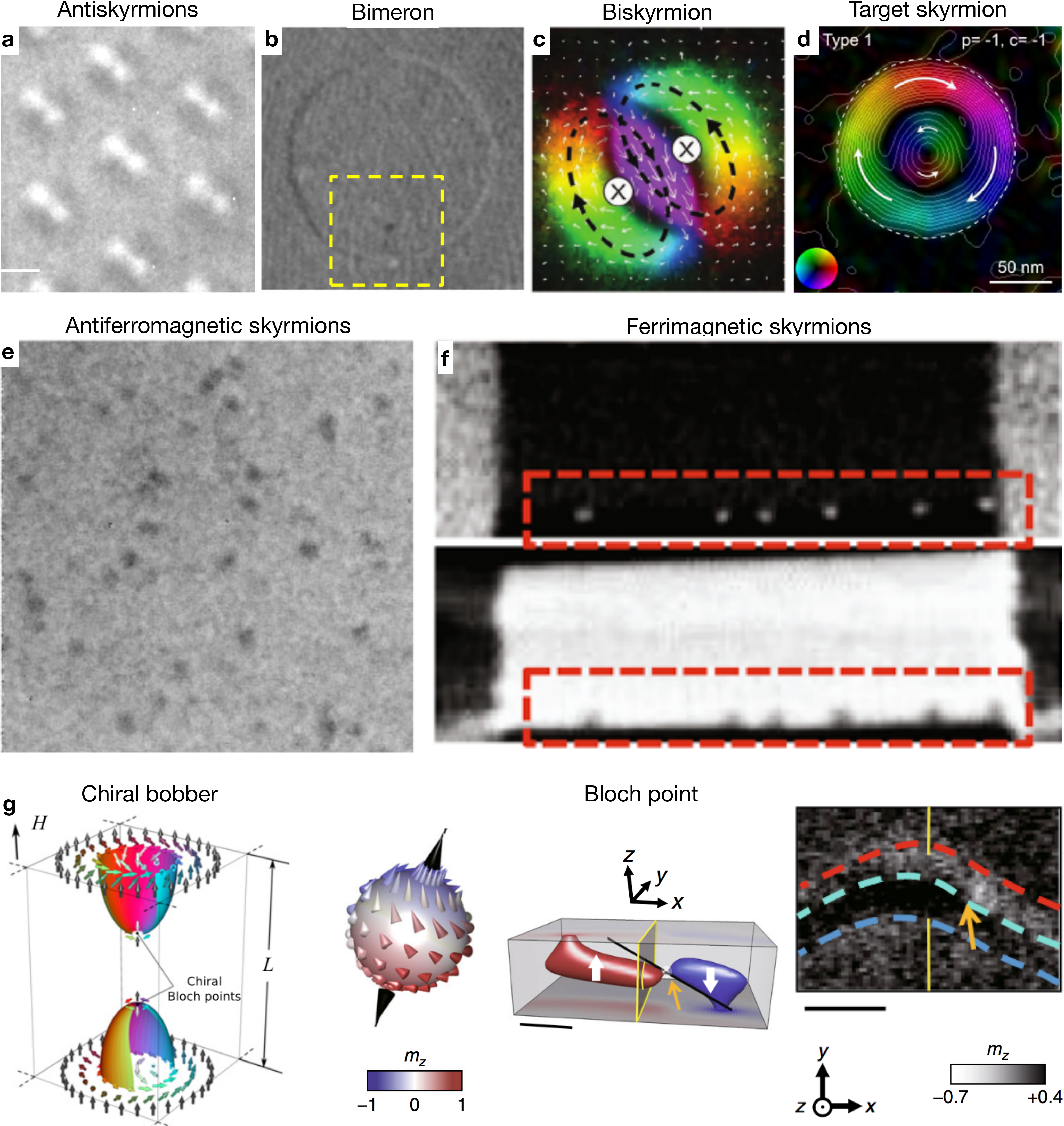}
  \caption{Several experimentally observed alternative magnetic nano-objects. (a) shows the LTEM contrast of antiskyrmions in the Heusler material Mn$_{1.4}$Pt$_{0.9}$Pd$_{0.1}$Sn (taken from~\cite{jena2020elliptical}). Dark and bright contrasts correspond to the different Bloch parts of the antiskyrmion. (b) A bimeron in a continuous Py film generated by local vortex imprinting from a Co disk measured by full-field magnetic transmission soft X-ray microscopy (MTXM) (taken from~\cite{gao2019creation}). In (c) the magnetic texture of a biskyrmion is shown in MnNiGa (taken from~\cite{li2019oriented}). (d) shows a target skyrmion observed by off-axis electron holography in nano-pillars of the B20 material FeGe (taken from~\cite{zheng2017direct}). It is resembling a skyrmionium. (e) shows the magneto-optical Kerr effect microscopy (MOKE) measurements of synthetic antiferromagnetic skyrmions in a multilayer stack (taken from~\cite{dohi2019formation}), which however has a small net moment. In (f) the scanning transmission X-ray microscopy (STXM) images of ferrimagnetic skyrmions in a GdFeCo film are shown (taken from~\cite{woo2017current}). (g) shows a schematic representation of two chiral bobbers (taken from~\cite{rybakov2016new}). While in~\cite{zheng2018experimental} chiral bobbers have been distinguished from skyrmion tubes by a lower contrast in a LTEM measurement, the right panels show a schematic representations and a magnetic transmission soft X-ray microscopy (MTXM) measurement of a Bloch point embedded in deformed magnetic vortex cores in
asymmetrically shaped Ni$_{80}$Fe$_{20}$ nanodisks (taken from \cite{im2019dynamics}). In this material the Bloch point can be resolved directly without the superposition of the corresponding skyrmion tube profile. All panels are taken (and have been rearranged) from publications (as indicated) that were published under a Creative Commons license.}
  \label{fig:experiments}
\end{figure}

In this section, we focus on skyrmions with an intermediate helicity~\cite{okubo2012multiple,garlow2019quantification} and antiskyrmions~\cite{nayak2017magnetic}. Due to their non-trivial topological charge, both objects exhibit a topological Hall effect and a skyrmion Hall effect in the STT scenario. However, in the SOT scenario the possibility to move these objects parallel to an applied current exists~\cite{huang2017stabilization,kim2018asymmetric}. In the following, we elaborate on this motion, the particles' stability and other non-trivial observations.

\subsubsection{Skyrmions with arbitrary helicity}

\begin{figure}[t!]
  \centering
  \includegraphics[width=\textwidth]{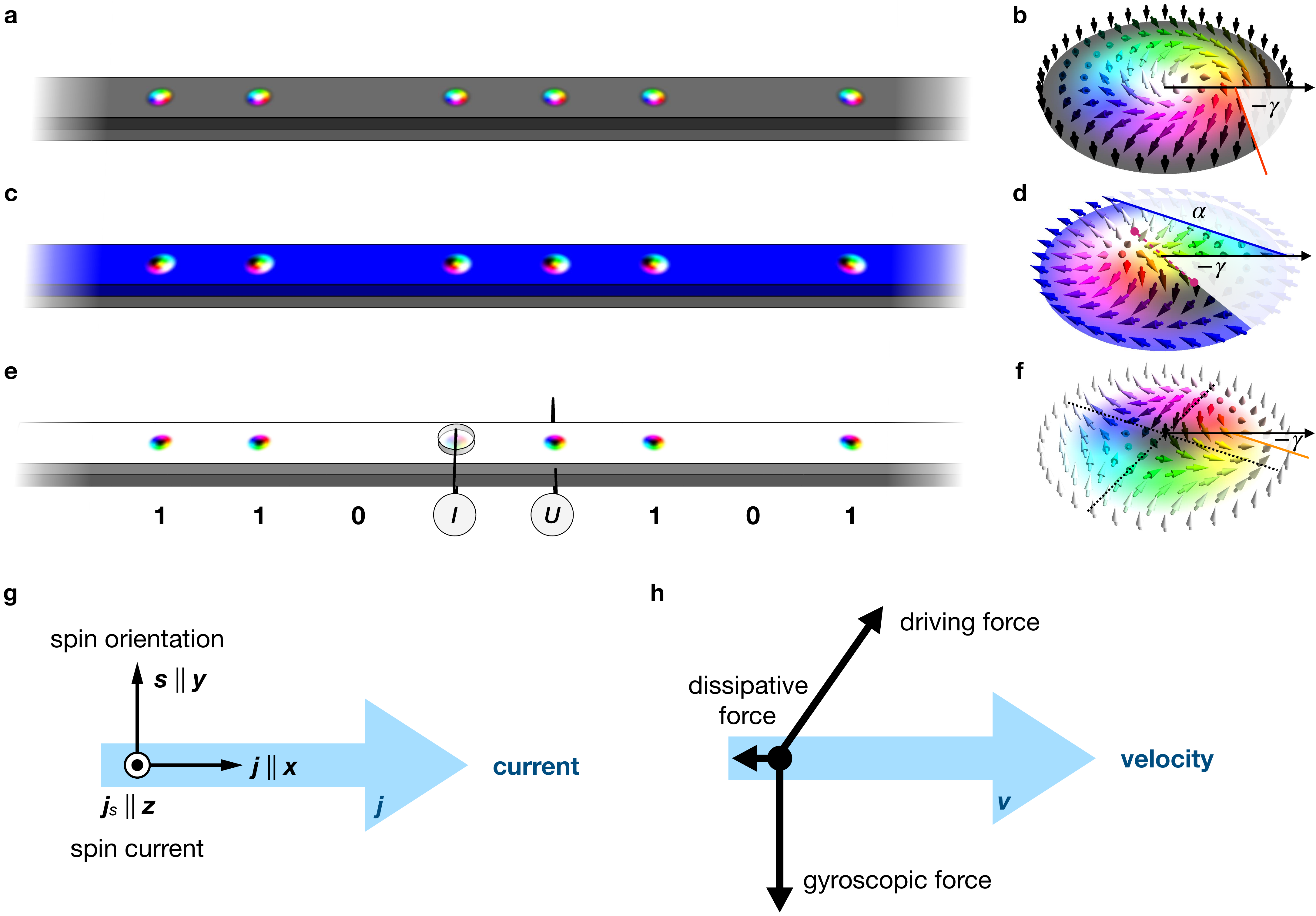}
  \caption{Straight-line motion of nano-objects in a racetrack. (a) Magnetic skyrmions move in the middle of a racetrack since the helicity of the skyrmion $\gamma$ has a specific value as indicated in (b). (c) Likewise, magnetic bimerons can move in an in-plane magnetized racetrack without skyrmion Hall effect. These objects are characterized by two angles $\alpha$ and $\gamma$, as indicated in (d). (e) Antiskyrmions can also move along a racetrack without skyrmion Hall effect if they are oriented at a certain angle $\gamma$, as indicated in (f). All of the presented objects have the same topological charge of $N_\mathrm{Sk}=1$ but the magnetic background is different in all cases. In (g) the applied currents are explained and in (h) the resulting forces are indicated. If specific types of skyrmions, bimerons or antiskyrmions are selected, the transverse component of the driving force from the injected spins compensates the gyroscopic force due to the topological charge. All figures are schematic.}
  \label{fig:racetrack}
\end{figure}

One main problem upon utilizing magnetic skyrmions in racetrack devices is the skyrmion Hall effect. A possible solution is the utilization of skyrmions with a helicity $\gamma$ different from that of N\'{e}el or Bloch skyrmions. In DMI systems this can in principle be achieved by considering a mix of interfacial and bulk DMI ~\cite{kim2018asymmetric} or by considering interfacial DMI in materials where the dipole-dipole interaction (favoring Bloch skyrmions) is considerable~\cite{buttner2018theory,everschor2018perspective}. Such objects have recently been observed using LTEM imaging~\cite{garlow2019quantification}.

In order to understand the skyrmion Hall effect, we analyze the tensors in the Thiele equation. For a positive polarity, skyrmions with an arbitrary helicity are characterized by a topological charge of $N_\mathrm{Sk}=+1$ leading to a gyroscopic vector of $\vec{G}=-4\pi\,\vec{e}_z$ independent of the skyrmion's helicity. This independence holds also for the dissipative tensor which is diagonal with only $D_{xx}=D_{yy}\neq 0$. 
However, the torque tensor depends on the helicity $\gamma$, which implies that a SOT, characterized by spins $\vec{s}\parallel \vec{y}$ that are injected from the perpendicular direction $z$, drives skyrmions with different helicities differently, resulting in different skyrmion Hall angles and trajectories. In general, the torque tensor of a skyrmion has the shape of a rotation matrix $\underline{R}_z$ around the $z$ axis
\begin{equation}
\underline{I}=\begin{pmatrix} 
\sin\gamma & \cos\gamma & 0\\
-\cos\gamma & \sin\gamma & 0\\
0 & 0 & 0
\end{pmatrix}=\underline{R}_z(\gamma+\pi/2).
\end{equation}
For a particular helicity $\gamma$ (Fig.~\ref{fig:racetrack}a,b), the transverse motion of a skyrmion due to the gyroscopic force is compensated by a component of the driving force (Fig.~\ref{fig:racetrack}g) so that the skyrmion moves along the applied electric current. Likewise, there exists a helicity for which the skyrmion moves perpendicular towards the edge. Due to the recent observation of these objects~\cite{garlow2019quantification}, a verification of the theoretical predictions is highly anticipated.

\subsubsection{Antiskyrmions}

Antiskyrmions are characterized by a vorticity of $m=-1$, i.\,e., the in-plane magnetization rotates oppositely to the position vector. Unlike skyrmions (irrespective of their helicity), these particles are not rotationally symmetric. For this reason, the helicity has a different meaning for antiskyrmions. It is not a global offset between the polar angle of the position vector $\phi$ and the polar angle of the magnetization $\Phi$, but distinguishes two axes along which the antiskyrmion has the profile of a N\'{e}el skyrmion with helicity $0$ and $\pi$, respectively (dashed lines in Fig.~\ref{fig:racetrack}f). The profile looks different along other lines. For example, along the two bisectrices of these two axes the antiskyrmion has the same texture as a Bloch skyrmion with helicity $\pm \gamma/2$, respectively. 

Geometrically, an antiskyrmion can be constructed from a skyrmion by rotating all spins of the skyrmion by 180$^\circ$ around a distinguished in-plane axis -- say $y$. In this case, the $x$ and $z$ components of the magnetization change sign. The resulting antiskyrmion still has the same topological charge as the skyrmions since the polarity and the vorticity both change their signs. Furthermore, the rotation argument allows to also identify the DMI necessary to stabilize antiskyrmions: two of the DMI vectors need to change their sign~\cite{Gungordu2016,huang2017stabilization}. This leads to the anisotropic DMI [Eq.~\eqref{eq:DMIaniso}] presented in an earlier section. Note again, that the DMI vectors are determined by the crystal symmetry: While N\'{e}el skyrmions arise at interfaces, where heavy metal atoms are located directly below the bond of two magnetic atoms, for antiskyrmions a layered system is required with heavy metal atoms above and below two different bonds. This is the case in some Heusler materials, and indeed antiskyrmion crystals have been observed in Mn$_{1.4}$Pt$_{0.9}$Pd$_{0.1}$Sn~\cite{nayak2017magnetic} and Mn$_{2}$Rh$_{0.95}$Ir$_{0.05}$Sn~\cite{jena2019observation} -- Heusler materials with $D_{2d}$ symmetry that exhibit this particular type of DMI~\cite{hoffmann2017antiskyrmions,huang2017stabilization}. The antiskyrmions in th\change{ese} material\change{s} have been shown to have long lifetimes at room temperature~\cite{Potkina2019}. Since an antiskyrmion can be understood to consist of Bloch and N\'{e}el parts, the observed LTEM contrast (Fig.~\ref{fig:experiments}a) is unique. It exhibits two spots of high intensity and two spots of low intensity, corresponding to the different Bloch parts. 

The change of the vorticity compared to skyrmions is highly relevant for the current-driven motion of antiskyrmions. The gyroscopic vector still points along $z$ and the dissipative tensor $\underline{D}$ is still symmetric. However, the $\underline{I}$ tensor changes
\begin{equation}
\underline{I}=\begin{pmatrix} 
-\sin\gamma & -\cos\gamma & 0\\
-\cos\gamma & \sin\gamma & 0\\
0 & 0 & 0
\end{pmatrix}.
\end{equation}
Like for the case of skyrmions with an arbitrary helicity, one can identify a particular helicity for which the skyrmion Hall effect is compensated for antiskyrmions. The important difference is that the helicity of antiskyrmions corresponds merely to a rotation in real space. Consequently, applying the current along a certain direction leads to an antiskyrmion motion parallel to the current, as presented in Ref.~\cite{huang2017stabilization}. This allows to think of a racetrack, where the D$_{2d}$ material is cut at an angle with respect to the high symmetry directions \cite{jena2020formation}. In a rotated coordinate system, where one axis points along the current direction, this leads to a rotation of the effective micromagnetic DMI vectors and generates a rotated antiskyrmion, i.\,e., an antiskyrmion with a certain $\gamma$. If it is cut under the correct angle, this racetrack allows for the desired straight-line motion of the bits without any transverse deflection (Fig.~\ref{fig:racetrack}e,f).

Another particularly interesting feature of antiskyrmion systems is that their aniso\-tro\-pic DMI is in conflict with the ubiquitous dipole-dipole interaction. While the first interaction allows only for antiskyrmions, the dipole-dipole interaction energetically favors Bloch skyrmions. The coexistence of both topologically distinct nano-objects has recently been observed by LTEM measurements~\cite{jena2020elliptical,peng2020controlled} and confirmed by micromagnetic simulations~\cite{jena2020elliptical}. Their distinct topological charges have later been confirmed by topological Hall measurements \cite{sivakumar2019double}. The DMI-stabilized antiskyrmions may show a square-shaped deformation due to the perturbative effect of the dipole-dipole interaction. The Bloch parts are increased and the N\'{e}el parts shrink in order to minimize the dipole-dipole energy~\cite{jena2020elliptical}, as predicted in Ref.~\cite{camosi2018micromagnetics}. The dipole-dipole-mediated Bloch skyrmions, that come in two flavors (helicities $\gamma=\pm\pi/2$), are elliptically deformed due to the anisotropic DMI~\cite{jena2020elliptical}. This finding allows to generalize the concept of the racetrack device using both, antiskyrmions and skyrmions, as the bits of information~\cite{jena2020formation,jena2020elliptical} (Fig.~\ref{fig:other}a). Such devices would be insensitive to diffusion or interactions between the nano-objects, since non-periodic bit sequences are unproblematic in this case. The coexistence of skyrmions and antiskyrmions has also been predicted by frustrated exchange interactions~\cite{okubo2012multiple} and for a very specific DMI tensor~\cite{hoffmann2017antiskyrmions}. However, these two cases remain to be observed experimentally.


\subsection{Combination of skyrmions or merons} \label{sec:combination}

In this section we present alternative magnetic quasiparticles that are formed as the combination of two or more skyrmions or merons. We begin with the bimeron~\cite{kharkov2017bound,gobel2018magnetic,gao2019creation}. It is the combination of two merons, which need to possess opposite vorticities in order to be geometrically compatible with each other and with the ferromagnetic background. Since also their polarities are mutually reversed, a bimeron is characterized by a topological charge of $N_\mathrm{Sk}=\pm 1$. It can therefore be considered a skyrmion in an in-plane magnetized material. Due to their missing rotational symmetry, bimerons are highly relevant for racetrack applications similar to antiskyrmions and intermediate skyrmion as presented in the last section.

Biskyrmions on the other hand, are formed by two partially overlapping skyrmions giving the new texture a topological charge of $N_\mathrm{Sk}=\pm 2$~\cite{yu2014biskyrmion,gobel2019forming}. In order to be compatible, this time the two Bloch skyrmions need to have opposite helicities instead of vorticities.

When instead two skyrmions with opposite polarities are combined, the new object has a vanishing topological charge. This makes skyrmioniums~\cite{zhang2016control,zhang2018real} and antiferromagnetic skyrmions~\cite{barker2016static,legrand2019room} the optimal candidates for spintronic applications. In the corresponding sections we discuss differences of these two objects, especially regarding the Hall effect of electrons.

Lastly, it is also worth mentioning that this list is non-exhaustive. For example, in synthetic antiferromagnets based on Co/Ru/Co films bimeronic topological defects (including both bi-vortices and bi-antivortices) living on domain walls were experimentally observed~\cite{Kolesnikov2018}. Furthermore, a pair of helix edges is considered the fundamental topological excitation in a helical phase~\cite{muller2017magnetic} and a 360$^\circ$ rotation of the internal magnetization in a domain wall can be called a domain wall skyrmion~\cite{cheng2019magnetic,li2020experimental}.

\subsubsection{Bimerons as in-plane skyrmions}

\begin{figure}[t!]
  \centering
  \includegraphics[width=1.0\textwidth]{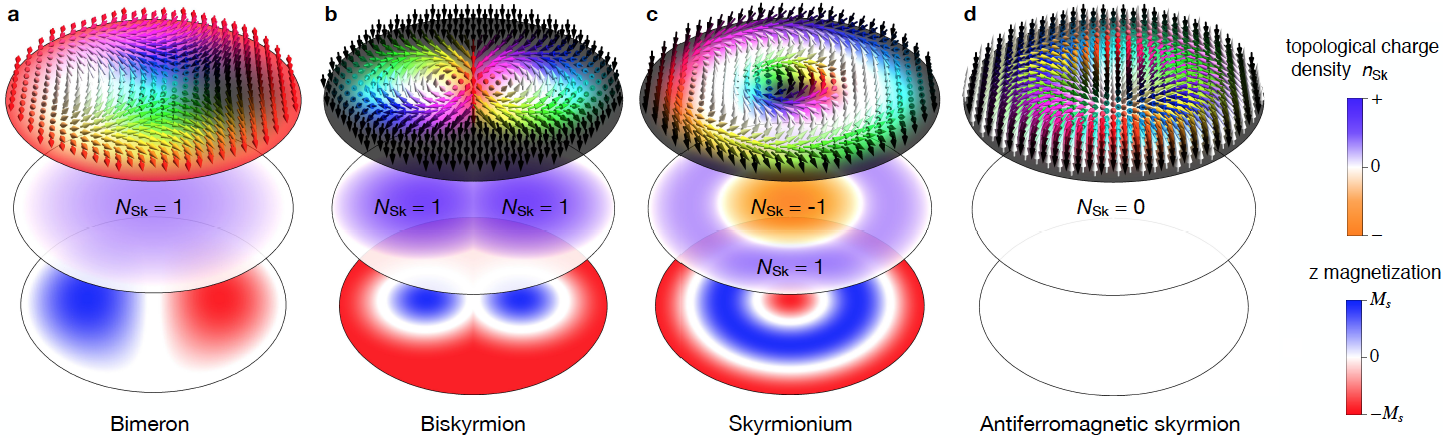}
  \caption{Texture, magnetization and topological charge density of selected nano-objects. In (a) a bimeron with $N_\mathrm{Sk}=1$ is shown (top panel). The middle panel shows the topological charge density that integrates to the topological charge as indicated. The bottom panel shows the out-of-plane magnetization. (b)-(d) show the biskyrmion, skyrmionium and antiferromagnetic skyrmion, respectively.}
  \label{fig:biskyrmion}
\end{figure}

The first object discussed in this category is the magnetic bimeron (Fig.~\ref{fig:biskyrmion}a). It was first predicted in 2017 as a bimeron crystal~\cite{kharkov2017bound} and shortly after also as an individual particle~\cite{gobel2018magnetic}. 

As the name indicates, a bimeron consists of two merons -- more precisely of a meron and an antimeron with mutually reversed out-of-plane magnetizations (bottom panel in Fig.~\ref{fig:biskyrmion}a), i.\,e. opposite polarities. This gives both subparticles the same topological charge of either $+1/2$ or $-1/2$ and the bimeron a topological charge of $\pm 1$ (middle panel in Fig.~\ref{fig:biskyrmion}a). In this sense, it can also be considered a skyrmionic excitation in an in-plane magnet~\cite{gobel2018magnetic}. This becomes even more apparent from the following transformation: Starting from a conventional magnetic skyrmion, a rotation of all magnetic moments around a common in-plane axis by 90$^\circ$ in magnetization space results in a bimeron texture. Such rotations leave the topological charge invariant. A bimeron can therefore ambivalently be understood as a meron-antimeron pair with opposite polarities or as a skyrmion in an in-plane magnet~\cite{gobel2018magnetic}.

Compared to the skyrmion, the bimeron is not characterized by an integer polarity, but the background magnetization can be rotated freely in the plane, mathematically speaking. This means that the class of bimerons has an additional continuous degree of freedom. A bimeron is characterized by two continuous angles instead of a discrete polarity and a continuous helicity. A possible characterization is that $\gamma$ defines the angle of the connecting line between the two merons' centers with respect to the $x$ axis and $\alpha$ defines the orientation of the net magnetization (parallel to the surrounding) with respect to the $x$ axis, as indicated in Fig.~\ref{fig:racetrack}e.

Up to now, only individual bimerons have been observed experimentally that were generated in a Py film via local vortex imprinting from a Co disk~\cite{gao2019creation} (Fig.~\ref{fig:experiments}b). Furthermore, bubbles have been observed in in-plane magnets possibly pointing towards the existence of topologically non-trivial spin textures in in-plane magnetized samples~\cite{chen2017out}. Moreover, short skyrmion tubes were observed in MnSi whose cross-section along the tube direction has the profile of a bimeron~\cite{meynell2017neutron}. However, the three-dimensional continuation of a bimeron would be a bimeron tube which is not realized there. The observed objects are more similar to skyrmions than to bimerons. Furthermore, intermediate states between skyrmions
and bimerons can exist in ferromagnets with in-plane anisotropy when the conventional interfacial DMI is present~\cite{leonov2017asymmetric}. These asymmetric skyrmions have finite in-plane and out-of-plane net magnetizations.

The reason why only a handful of potential bimeron systems have been identified experimentally, may originate in the required DMI. One example generating one particular type of bimeron is given by~\cite{gobel2018magnetic}
\begin{equation}
H_\mathrm{bimeron}=\int \tilde{D}\big(m_z\frac{\partial m_x}{\partial x}-m_x\frac{\partial m_z}{\partial x}+m_x\frac{\partial m_y}{\partial y}-m_y\frac{\partial m_x}{\partial y}\big)\mathrm{d}^3r.
\end{equation}
Since the DMI vectors are rotated around an in-plane axis compared to the interfacial DMI, the same must also apply to the texture. In this rather complicated setup, the DMI vectors along one direction are oriented in-plane, just like for interfacial and bulk DMI. However, along the other bond direction, the DMI vectors point out-of-plane, meaning that the inversion symmetry has to be broken in a particular way (an example is presented in Ref.~\cite{gobel2018magnetic}). The required crystal structure lacks major symmetries, which is why most typical materials do not allow for a formation of bimerons. However, a similar DMI setup was recently predicted by first-principles calculations of Janus monolayers of the chromium trihalides Cr(I,Cl)$_3$ and Cr(I,Br)$_3$ for which bimerons have been stabilized in simulations \cite{xu2020topological}. Furthermore, it has been predicted that bimerons can be stabilized much more easily by frustrated exchange interactions. In contrast to the DMI, frustrated exchange interactions are achiral allowing to stabilize skyrmions, antiskyrmions, bimerons and even other objects likewise~\cite{okubo2012multiple,kharkov2017bound,gobel2018magnetic,Zhang2020bimeron}. The only requirement for bimerons is that the external magnetic field has to be applied in the plane. Until now, topologically non-trivial spin textures with these properties stabilized by frustrated exchange interactions have never been observed. However, once this is the case, a rotation of the magnetic field will result in bimeron formation if the anisotropy is weak (or even better in the plane) and if the DMI is negligible. Without one of the two mechanisms bimerons can only exist as transition states that are unstable~\cite{komineas2007rotating,zhang2015magnetic,heo2016switching}.

The observation of bimerons is highly desirable due to their special emergent electrodynamics. Similar to what has been presented for antiskyrmions and skyrmions with an arbitrary helicity (see last section), bimerons have a reduced symmetry compared to Bloch or N\'{e}el skyrmions. While the dissipative tensor and the gyroscopic vector are the same as for skyrmions and antiskyrmion, the global spin rotation manifests in the torque tensor, which is not antisymmetric anymore~\cite{gobel2018magnetic}. Following the same argumentation as for the antiskyrmion, there exists a specific current direction for which bimerons move parallel to the current (specific $\gamma$, $\alpha$ combinations), implying that specific bimerons can be used in racetracks as bits that move without skyrmion Hall effect~\cite{vk}, as shown in Fig.~\ref{fig:racetrack}c.
In the STT scenario on the other hand, bimerons will always experience the skyrmion Hall effect, since the adiabatic torque transforms like the texture. Moreover, it was shown that if a suitable magnetic field gradient is introduced, an alternating magnetic field can also move the bimeron in such a way that the skyrmion Hall effect does not show up \cite{shen2020dynamics}.

\begin{figure}[t!]
  \centering
  \includegraphics[width=.9\textwidth]{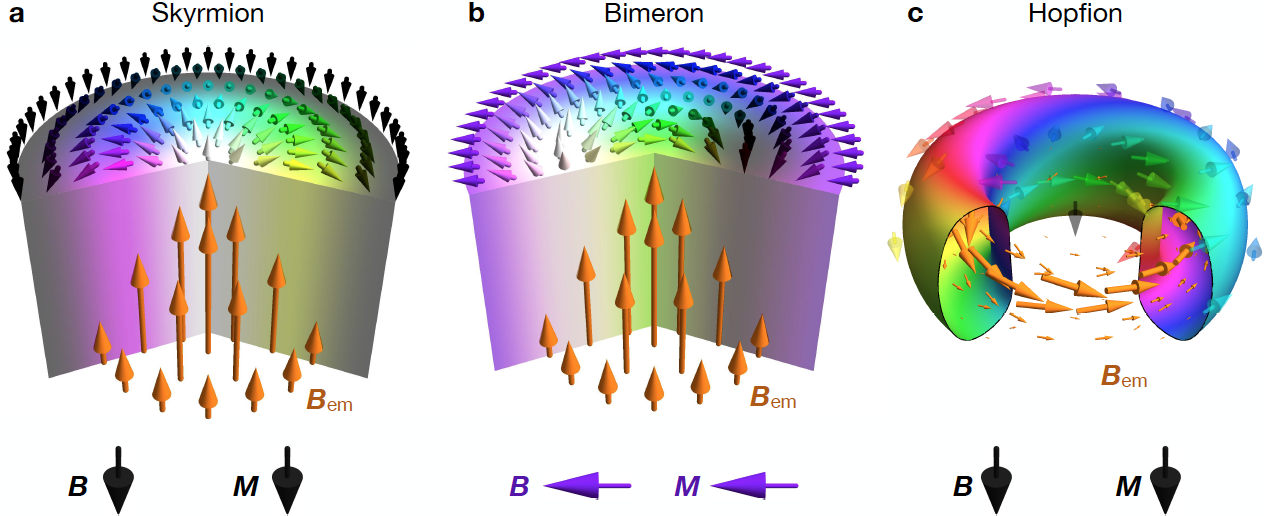}
  \caption{Electronic transport behavior of different nano-objects. (a) Skyrmion tube with emergent field $\vec{B}_\mathrm{em}$ along the tube direction ($z$; orange arrows). The net magnetization $\vec{M}$ and the stabilizing field $\vec{B}$ are (anti-) parallel giving rise to only $xy$ and $yx$ elements of the topological, anomalous and ordinary Hall resistivity. (b) Bimeron tube. The emergent field is not parallel to the net magnetization and the required stabilizing field meaning that the different contributions to the Hall effect appear in different tensor elements. (c) Magnetic hopfion. Again, the emergent field (locally mainly in-plane and globally compensated) is not parallel to the net magnetization and the stabilizing field, as indicated.}
  \label{fig:differentemergent}
\end{figure}

Furthermore, the bimeron is unique in terms of the topological Hall effect. 
In Fig. \ref{fig:differentemergent}b a bimeron tube is shown. Its emergent field $\vec{B}_\mathrm{em}$ is along the tube direction (orange arrows along $z$), since the texture only changes in the $xy$ plane. Therefore, the topological Hall effect manifests itself only in the $xy$ (and $yx$) tensor element of the resistivity, as for a magnetic skyrmion tube. However, since a bimeron is an excitation in an in-plane magnet, the net magnetization $\vec{M}$ is in the plane as well (purple arrow). Especially, the out-of-plane component is zero, meaning that no anomalous Hall effect enters the $xy$ resistivity tensor element -- at least by conventional means, as introduced in Sec. \ref{sec:emergent} \change{[Eq. \eqref{eq:hall}]. However, as discussed earlier, the anomalous Hall effect has even been observed in coplanar systems without a net magnetization \cite{nakatsuji2015large,nayak2016large}, depending on the system's symmetry. Therefore, for bimeron hosts with certain crystal symmetries, it may still be possible that an anomalous Hall effect arises in the $xy$ resistivity tensor element even though $M_z=0$}. In a topological Hall measurement, the stabilizing field (typically along $z$ for the skyrmion) is swept. For the bimeron this field has to point in-plane (purple arrow $\vec{B}$) meaning that also no conventional Hall effect is measurable in $\rho_{xy}^\mathrm{HE}$.
For this reason, bimerons are expected to allow for a pure measurement of the topological Hall effect upon variation of the in-plane field, establishing the optimal playground to investigate topological Hall physics~\cite{gobel2018magnetic}. This is a unique property of bimerons, different from skyrmion\change{s} or other nano-objects in out-of-plane magnetized materials for which the three contributions to the Hall effect are superimposed in the $xy$ tensor element of the resistivity, since the emergent field $\vec{B}_\mathrm{em}$, the net magnetization $\vec{M}$ and the stabilizing field $\vec{B}$ are all (anti-)parallel to each other (Fig. \ref{fig:differentemergent}a). By analogy with this discussion, the orbital magnetization is occurring along the bimeron tube direction. It is therefore perpendicular to the net moment of the spin texture and could be detected and addressed individually~\cite{gobel2018magnetic}.

As a final remark, we would like to mention that sometimes the term bimeron is also used for an elongated skyrmion (or short helix segments), where both ends of the skyrmion have topological charges of $\pm1/2$ and the center part is neutral~\cite{ezawa2011compact,du2013field,silva2014emergence,iakovlev2018bimeron}. This object is actually more similar to a skyrmion than to the bimeron in the sense of what has been discussed here.

\subsubsection{Biskyrmions}
The term biskyrmion is commonly used for two partially overlapping skyrmions (Fig.~\ref{fig:biskyrmion}b) first observed in 2014~\cite{yu2014biskyrmion}. In order to be geometrically compatible (the magnetization between the two skyrmions must be steady), both skyrmions need to possess a helicity difference of $\pi$, meaning that the in-plane magnetizations of both skyrmions are mutually reversed. 

Following from our explanations about the possible stabilizing mechanisms of skyrmions in Sec.~\ref{sec:stability}, biskyrmions can hardly be stable when the DMI is large. This chiral interaction would prefer one particular skyrmion helicity leading to a discontinuous magnetization when two skyrmions would partially overlap (in that case, even two attractive skyrmions would always remain separated by a ferromagnetic area, like in Ref.~\cite{du2018interaction}). On the contrary, the dipole-dipole interaction is achiral: it equally favors both types of Bloch skyrmions with helicities of $\gamma=\pm\pi/2$, respectively~\cite{malozemoff2016magnetic,gobel2019forming}. When these two skyrmions are partially superimposed, the in-plane magnetizations between the skyrmions' centers point along the same direction, meaning that they are geometrically compatible. And indeed, it was found that already the short-range approximation of the dipole-dipole interaction leads to an attraction between two Bloch skyrmions with opposite helicities allowing to form magnetic biskyrmions~\cite{gobel2019forming}. Besides the stabilization of individual magnetic biskyrmions, biskyrmion lattices have been stabilized by this mechanism in micromagnetic simulations as well~\cite{capic2019stability,capic2019biskyrmions}.

Even before the theoretical prediction, biskyrmion lattices have been observed experimentally (Fig.~\ref{fig:experiments}c) in centrosymmetric materials~\cite{yu2014biskyrmion,wang2016centrosymmetric,peng2017real, zuo2018direct, peng2018multiple} like the layered manganite La$_{2-2x}$Sr$_{1+2x}$Mn$_2$O$_7$. In these materials the DMI is absent by symmetry, in agreement with the theoretically established stabilizing mechanism. Another possible origin for biskyrmion formation may be frustrated exchange interactions, as predicted in Ref.~\cite{zhang2017skyrmion}.

Regarding their emergent electrodynamics, biskyrmions behave similar to conventional skyrmions. Their topological charge of $N_\mathrm{Sk}=\pm 2$ (middle panel in Fig.~\ref{fig:biskyrmion}b) leads to the emergence of a topological Hall effect and a skyrmion Hall effect in the STT scenario. The current-driven motion under SOT may turn out problematic, since the opposite helicities of the two Bloch skyrmions lead to sign reversed $\underline{I}$ tensors for the two subskyrmions. Consequently, they are driven along different directions eventually leading to the destruction of the biskyrmion. However, the lack of rotational symmetry -- even in the $m_z$ component (bottom panel in Fig.~\ref{fig:biskyrmion}b) -- allows to utilize a rotation of biskyrmion for spintronic devices. As has been shown by micromagnetic simulations on a square lattice, the principle axis of a biskyrmion aligns with high-symmetry directions of the lattice (the second-nearest neighbor directions in this case), allowing in principle to switch between different metastable configurations~\cite{gobel2019forming}. 

As a closing remark, it is worth mentioning that the initial observations of bi\-skyr\-mi\-on crystals by LTEM imaging are under debate right now. In two recent publications~\cite{loudon2019images,yao2019magnetic} the authors showed that the unique Lorentz contrast can also arise due to a tilting effect. Tubes of topologically-trivial bubbles appear as skyrmion pairs with reversed in-plane magnetizations when viewed under an angle. Still, as explained, the stability of biskyrmions has recently been explained theoretically~\cite{gobel2019forming,capic2019stability,capic2019biskyrmions}, so the observed textures may indeed be biskyrmions. Alternative experimental techniques have to be utilized to dispel remaining doubts on the existence of magnetic biskyrmions.

\subsubsection{Skyrmioniums}
Up to now, we have discussed fundamental and composite magnetic objects with a finite topological charge. If however two skyrmions with opposite topological charges are combined to a new particle, the topological charge vanishes. If both objects would have the same polarity, this would require the combination of objects with opposite vorticities, e.\,g. a skyrmion and an antiskyrmion. Although these objects have been found to coexist in D$_{2d}$ materials~\cite{jena2020elliptical}, a combination of both has not yet been observed. 

If however both objects have the same vorticities but opposite polarities, they can even be stabilized by bulk or interfacial DMI -- the interaction that is typically dominating in skyrmion hosts. In fact, it has been found that under certain conditions an increase of the DMI constant leads to the formation of a skyrmion inside of another skyrmion with mutually reversed spins~\cite{zhang2016control} (Fig.~\ref{fig:biskyrmion}c). This object is called skyrmionium or $2\pi$-skyrmion~\cite{bogdanov1999stability,beg2015ground, finazzi2013laser,zheng2017direct,zhang2018real,hagemeister2018controlled, kolesnikov2018skyrmionium,li2018dynamics,shen2018motion, pylypovskyi2018chiral}, since the azimuthal angle rotates by $2\pi$ (instead of $\pi$) when going from the object's center to the confinement. 
The outer skyrmion has the shape of a ring and is characterized by the opposite topological charge compared to the inner skyrmion (middle panel in Fig.~\ref{fig:biskyrmion}c). 

Skyrmioniums have recently been observed in a thin ferromagnetic film on top of a topological insulator~\cite{zhang2018real} and have been created by laser pulses~\cite{finazzi2013laser}. Furthermore, their generation has been predicted by the perpendicular injection of spin currents~\cite{zhang2016control,kolesnikov2018skyrmionium,gobel2019electrical} or via alternating the out-of-plane orientation of an external magnetic field~\cite{hagemeister2018controlled}. Moreover, the static and dynamic properties of an isolated nanoscale skyrmionium were recently reported in a frustrated magnetic monolayer, where the skyrmionium is stabilized by competing exchange interactions. The skyrmionium in frustrated systems may have a very small size of 10 nm, which can be further reduced by tuning the perpendicular magnetic anisotropy or the magnetic field~\cite{xia2020skyrmionium}.

The concept of the 2$\pi$-skyrmion can be extended to $k\pi$-skyrmions, theoretically stable for certain parameters when the DMI is increased even further. Experimentally, such objects have been seen in confined geometries like nanodisks~\cite{zheng2017direct,cortes2019nanoscale}. There, the texture is mainly stabilized by the confinement and the value of $k$ is determined by the disk radius. Such objects are also called target skyrmions due to the peculiar profile of the magnetization's $z$ component (Fig.~\ref{fig:biskyrmion}c)~\change{\cite{leonov2014target}}. In these disks the value of $k$ is often not integer since the magnetic moments are not pointing perfectly out-of-plane at the confinement (Fig.~\ref{fig:experiments}d). Another texture related to $k\pi$-skyrmions are so called skyrmion bags~\cite{foster2018composite,rybakov2019chiral} where multiple skyrmions (instead of just one) are positioned next to each other inside of a skyrmion ring.

Due to the vanishing topological charge and the observation in experiments, magnetic skyrmioniums seem to be highly relevant for spintronic applications in the future. Their emergent electrodynamics will be discussed in the following. The trivial real-space topology on a global level leads to the absence of a skyrmion Hall effect~\cite{zhang2016control}. For this reason, N\'{e}el-type skyrmioniums move without skyrmion Hall effect even in the SOT scenario. Still, the finite topological charges of the two subskyrmions become apparent: the skyrmionium deforms upon motion along the track. The inner skyrmion pushes towards the racetrack edge but is confined by the outer ring skyrmion that pushes transversally along the other direction. Both transverse motions compensate each other on a global level but still this effect constitutes a barrier for the maximum driving current~\cite{zhang2016control,gobel2019electrical}. When the current is too large, the transverse forces become too strong and the skyrmionium unzips~\cite{zhang2016control}. Consequently, both skyrmionium parts annihilate. Summarizing, the absence of the skyrmion Hall effect solves an important problem: a skyrmionium moves parallel to an applied current directly in the middle of a racetrack avoiding pinning at the edges. Still, the skyrmion Hall effect of the subsystems is problematic as the current density cannot be increased much higher than that of a conventional skyrmion system.

For the topological Hall effect one has to distinguish between global and local arguments as well. On a global level (measuring the Hall voltage of the charge accumulation on a mesoscopic length scale), the Hall effect is absent due to the vanishing topological charge. If, however, the detection is performed locally on the length scale of the skyrmionium size, one can find signatures of both subsystems. In Ref.~\cite{gobel2019electrical} it has been shown that the topological Hall effect of skyrmioniums in a racetrack exhibits a unique triple-peak feature originating from the alternating topological charges of the ring- and center-skyrmions. This effect can therefore be used to detect skyrmioniums electrically, in a similar way as presented for skyrmions~\cite{maccariello2018electrical,hamamoto2016purely}.

\subsubsection{Antiferromagnetic skyrmions}

The dynamic properties and the stability of single antiferromagnetic skyrmions have first been predicted in antiferromagnets~\cite{barker2016static,zhang2016antiferromagnetic} and synthetic antiferromagnetic bilayers~\cite{zhang2016magnetic}. Shortly after, they have been extended to two-sublattice antiferromagnetic skyrmion crystals~\cite{gobel2017afmskx}.
Like skyrmioniums, antiferromagnetic skyrmions can be understood as the combination of two skyrmions with mutually reversed spins. Therefore, they are characterized by a vanishing topological charge. However, in the present case, the subskyrmions are not spatially separated but intertwined. For this reason, the magnetization density vanishes locally and the N\'eel order parameter, the main order parameter for antiferromagnets, can be considered instead of the magnetization. Calculating the topological charge with this parameter gives $\pm 1$. Thus, the antiferromagnetic skyrmions are still skyrmions from the viewpoint of topology, but exhibit different dynamics compared to ferromagnetic skyrmions. These antiferromagnetic dynamics can also be described by the Thiele equation but by using the N\'eel order parameter~\cite{Tveten2013}. 

Due to the compensated topological charge for the magnetization, an antiferromagnetic skyrmion moves without skyrmion Hall effect~\cite{barker2016static,zhang2016magnetic}, similar to what has been discussed for the skyrmionium. However, the two subsystems are coupled much stronger and do not allow for a deformation due to a pairwise opposing transverse motion, as was the case for the skyrmionium. Typical for antiferromagnetic spin textures, the antiferromagnetic skyrmions can be propelled much faster by currents compared to conventional skyrmions. Velocities on the scale of kilometers per second have been simulated~\cite{barker2016static,zhang2016magnetic,jin2016dynamics}.  This makes antiferromagnetic skyrmions the ideal carriers of information for data storage devices. Additionally, it was theoretically shown that the diffusion constant~\cite{barker2016static} of antiferromagnetic skyrmions is high in systems with small damping unlike for their ferromagnetic counterparts, thus showing a potential in driving them with temperature gradients.
Furthermore, they do not exhibit stray fields, which potentially allows for a denser stacking of quasi one-dimensional racetracks upon building the three-dimensional storage device. 

In terms of the required DMI, the stabilization of antiferromagnetic skyrmions is not problematic~\cite{barker2016static,Bessarab2019}. \change{If one considers the  type of skyrmion that is energetically preferred by the DMI in a system (determined by the symmetry), a skyrmion with mutually reversed spins (i.\,e. a skyrmion with opposite polarity and a helicity difference of $\pi$) is energetically stable as well.} Furthermore, both subskyrmions need to be coupled antiferromagnetically quite strongly, to accomplish the antiparallel alignment of the corresponding magnetic moments.
And indeed, very recently, the bilayer-type antiferromagnetic skyrmions have been observed in synthetic antiferromagnets~\cite{legrand2019room,dohi2019formation} (Fig.~\ref{fig:experiments}e) at room temperature. In Ref.~\cite{legrand2019room} the small stray fields resulting from the bilayer setup have been detected using magnetic force microscopy (MFM). In Ref.~\cite{dohi2019formation} the authors explain a method to prepare synthetic antiferromagnets with a tunable net moment. While they can achieve a completely compensated system, they deliberately prepared also a system with a small net moment to be able to perform magneto-optical Kerr effect (MOKE) measurements. 

For antiferromagnetic skyrmion-based logic~\cite{Liang2020logic} or racetrack~\cite{barker2016static} applications, a controlled generation process is needed. Conventional skyrmions have for example been generated by directed, deterministic approaches such as spin torques or magnetic fields (see reviews~\cite{nagaosa2013topological,kang2016skyrmion,wiesendanger2016nanoscale, garst2017collective,finocchio2016magnetic,fert2017magnetic, jiang2017skyrmions,everschor2018perspective}). These approaches are, however, difficult to utilize for antiferromagnetic skyrmions, since all vectorial quantities would have to act either only on one of the two subskyrmions, so that the other generates automatically due to the strong antiferromagnetic coupling, or they would have to act on both subskyrmions with an opposite sign. Both \change{approaches are} hardly feasible, since magnetic fields for example cannot change their sign on the length scale of the lattice constant. Stochastic processes, like the generation of nano-objects at defects or from the confinement (as shown for conventional skyrmions~\cite{romming2013writing,jiang2015blowing}), appear to be more advantageous in this regard. The stabilization of antiferromagnetic skyrmions becomes even more challenging when an antiferromagnetic skyrmion crystal shall be stabilized. By analogy with skyrmion crystals, a stabilizing magnetic field is inevitable. It has to be oriented along $\pm z$ for the two subsystems, respectively. This problem may be circumvented by growing a potential antiferromagnetic skyrmion host on top of a collinear antiferromagnet with the same crystal structure at the interface. By doing so, a staggered magnetic field is mimicked by the exchange interaction at the interface~\cite{gobel2017afmskx}.

For antiferromagnetic skyrmions in a single layer (not the case of a synthetic antiferromagnet bilayer) the detection is another problem. Both, the magnetization and topological charge density of the magnetization, are compensated globally and locally (middle and bottom panels in Fig.~\ref{fig:biskyrmion}d). These antiferromagnetic skyrmions would therefore appear invisible for the real-space techniques, such as MFM or LTEM. Furthermore, anomalous and topological Hall signatures do not exist. Luckily, a different hallmark has been predicted: the topological spin Hall effect~\cite{buhl2017topological,gobel2017afmskx,akosa2018theory}.
The resulting signal is the analogue of the conventional spin Hall effect, but originates in the non-collinearity of the spin texture.
The topological spin Hall effect can most \change{intuitively} be comprehended if one assumes two electronically uncoupled subskyrmions: Due to the opposite spin alignment, the emergent fields of the two subskyrmions are oriented oppositely. This leads to a transverse deflection of the electrons in opposite directions. The spins of the two species of electrons align with their respective texture and can therefore be considered as `spin up' and `spin down' states, again due to the opposite spin alignment. \change{However, since the sublattices are actually coupled, the above definition of an emergent magnetic field becomes problematic. Furthermore, to calculate the spin Hall conductivity from the reciprocal space properties, a non-Abelian formulation has to be considered to account for the sublattice-degenerate bands. Likewise, the spin-polarization of an electron propagating in an antiferromagnetic skyrmion does not align completely with the texture anymore. A spin-dependent orbital motion of the conduction electrons becomes relevant \cite{cheng2012electron,gomonay2015berry}. Still, a topological version of the spin Hall effect arises.}

Summarizing, one can have a positive feeling about the utilization of antiferromagnetic skyrmions in spintronic devices in the future. Synthetic antiferromagnetic skyrmions have been observed and, just recently, the current-driven motion of synthetic antiferromagnetic skyrmions has been realized \cite{dohi2019formation}. Furthermore, single layer antiferromagnetic skyrmions have been predicted. The topological spin Hall effect may play an essential role for observing these objects despite completely compensated magnetizations, stray fields and topological charge densities of the magnetization. Moreover, even in antiferromagnetic insulators the skyrmions were predicted and can potentially be moved by an electrically created anisotropy gradient~\cite{Shen2018afm} or by thermal gradients. 

As a closing remark, we would like to mention that the idea to combine skyrmions on different sublattices has been generalized in several works. Three skyrmion crystals can for example be intertwined as stabilized by Monte Carlo simulations~\cite{rosales2015three,diaz2019topological}. However, these objects do not exhibit the advantages of antiferromagnetic skyrmions because the topological charge for the magnetization is finite in this case.

\subsubsection{Ferrimagnetic skyrmions}

Signatures of the favorable emergent electrodynamics of antiferromagnetic skyr\-mi\-ons have also been seen for a similar object: the ferrimagnetic skyrmion~\cite{woo2017current,kim2017self} (Fig.~\ref{fig:particles}g). It consists of two coupled subskyrmions with mutually reversed spins, similar to the antiferromagnetic skyrmion. 

However, the magnetic moments have different magnitudes on the two sublattices leading to an uncompensated magnetization, which allowed for the detection of ferrimagnetic skyrmions in GdFeCo films by X-ray imaging~\cite{woo2017current} (Fig.~\ref{fig:experiments}f). When these objects are driven by spin currents, there exists a critical temperature at which the skyrmion Hall effect is absent~\cite{kim2017self}. At this temperature, the angular momentum is compensated, even though the magnetization is not, due to different gyromagnetic ratios for the magnetic moments in the different sublattices~\cite{kim2017self}. Experimentally, this complete compensation of the skyrmion Hall effect still lacks observation, but a reduced skyrmion Hall angle of $\theta_\mathrm{Sk}=20^\circ$ has been observed at room temperature~\cite{woo2017current}. Moreover, it was recently experimentally observed that in ferrimagnetic insulators near the compensation temperature domain walls driven by SOT can move at speeds reaching 6 km/s~\cite{Zhou2019compensated}, thus hinting that the same speeds can be achieved as well by ferrimagnetic skyrmions in the future. 

Compared to antiferromagnetic skyrmions, ferrimagnetic skyrmions bring about the advantage of a convenient detection and the possibility to address them due to their uncompensated magnetization. Concerning their emergent electrodynamics they promise similar advantages as antiferromagnetic skyrmions. However, a straight-line motion along a driving current is only expected to work at a particular (angular momentum compensation) temperature, which limits the spintronics applicability.

\begin{figure}[t!]
  \centering
  \includegraphics[width=\textwidth]{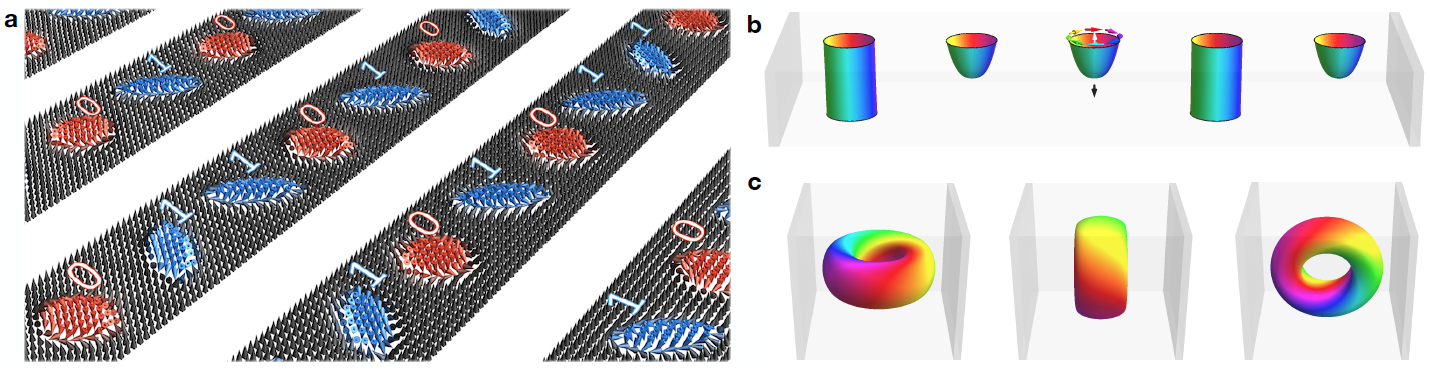}
  \caption{Further storage concepts. (a) In Heusler materials the anisotropic DMI stabilizes antiskyrmions (red) and elliptically deformed skyrmions (blue). These coexisting types of topologically distinct objects may encode data as '0' and '1' bits, respectively. The figure is taken from the press release \cite{press1} based on publication~\cite{jena2020elliptical}, copyright: B{\"o}rge G{\"o}bel, Martin-Luther-Universit{\"a}t Halle-Wittenberg. (b) Chiral bobbers can coexist with regular skyrmion tubes allowing for a similar type of device as shown in~\cite{zheng2018experimental}. (c) Magnetic hopfions as innately three-dimensional solitons can also be used as carriers of information in racetrack devices. Since they are anisotropic, in principle, they can be tilted~\cite{gobel2020topological,liu2020three} and potentially be switched to encode data.}
  \label{fig:other}
\end{figure}


\subsection{Three-dimensional objects} \label{sec:3d}
In the last main section we present three-dimensional solitons and extensions of skyrmions. First of all, we want to mention the trivial case: two-dimensional skyrmions, antiskyrmions, bimerons and other objects are extended as tubes along the third dimension. Still, these tubes can show interesting interactions and can even have a varying helicity along the tube~\cite{legrand2018hybrid}. However, what is even more interesting, is when these tubes interact, merge, begin, end or form closed loops, leading to the occurrence of Bloch points, chiral bobbers or hopfions.

First, we discuss Bloch points (the hedgehog in Fig.~\ref{fig:topology}b, also shown in an experimental measurement in Fig.~\ref{fig:experiments}g). These are the fundamental objects in three-dimensions: Around a singularity, where the magnetization is not well defined, the magnetic texture points into all directions of three-dimensional space. Closely related to these objects are chiral bobbers~\cite{rybakov2015new,zheng2018experimental} (Fig.~\ref{fig:particles}j). These are skyrmion tubes that end in a Bloch point. Interestingly, the chiral bobbers can appear in the same sample as skyrmion tubes. Both objects can be well distinguished experimentally, allowing to think of an improved racetrack storage device with both objects as the bits of information~\cite{zheng2018experimental} (Fig.~\ref{fig:other}b).
On the other hand, skyrmion tubes can also bend and form closed loops called hopfions~\cite{liu2018binding} (Fig.~\ref{fig:particles}l). These objects are unique from a fundamental point of view since they are characterized by a second topological invariant, the Hopf number. 

\subsubsection{Bloch points}

As explained, the trivial continuation of skyrmions along the third dimension are skyrmion tubes. In MnSi for example, they penetrate the whole sample. However, materials are not always perfect, so it can happen that two skyrmion tubes merge to one~\cite{wild2017entropy,birch2019increased,kagawa2017current}. Similarly, a single skyrmion tube can end in or begin from a Bloch point~\cite{wild2017entropy,kagawa2017current,birch2019increased, koshibae2019dynamics,birch2020real} -- a singularity in the local magnetization.

In three-dimensions the emergence of singularities is unavoidable for such textures: A change of topological charge in the cross section (from $N_\mathrm{Sk}=\pm 1$ in the skyrmion tube to $N_\mathrm{Sk}=0$ in the uniform ferromagnet) has to be condensed to a single point, since a topological charge cannot change continuously. While the corresponding singularity is a problem in a continuous model, the discreteness of magnetic moment circumvents this apparent problem~\change{\cite{doring1968point}}. Mathematically speaking, the singularity is located between the lattice sites. Around the singularity the magnetization points in all possible directions in three-dimensional space.

An isolated Bloch point has been imaged in asymmetrically shaped Ni$_{80}$Fe$_{20}$ nanodisks by magnetic transmission soft X-ray microscopy (MTXM) \cite{im2019dynamics} (Fig. \ref{fig:experiments}g). Furthermore, in Ref. \cite{birch2020real} scanning transmission X-ray microscopy (STXM) has been used to observe Bloch points at the end of skyrmion tubes in FeGe.

Besides considering skyrmion tubes starting from or ending in single Bloch points, a periodic lattice of Bloch and anti-Bloch points can be present, like in MnGe (Fig. \ref{fig:lattice}d). In MnSi$_{1-x}$Ge$_x$ this lattice exists in different types: either characterized by three $\vec{Q}$ vectors along the pairwise perpendicular $x$, $y$ and $z$ directions, or by four $\vec{Q}$ vectors spanning a tetrahedron \cite{fujishiro2019topological}. It has been shown that a conventional skyrmion tube lattice transitions to a 4$\vec{Q}$-hedgehog lattice upon increasing the Ge over Si ratio. When this ratio is further increased, the hedgehog lattice deforms and is now characterized by three $\vec{Q}$ vectors as in Fig. \ref{fig:lattice}d \cite{fujishiro2019topological}. A 3$\vec{Q}$-hedgehog lattice has been observed by LTEM in MnGe \cite{tanigaki2015real}. Its existence may be related to different types of magnetic interactions: In Ref. \cite{grytsiuk2020topological} it was shown that the typical Heisenberg interaction and DMI are able to explain the stability of quasi-two-dimensional skyrmions lattices in FeGe, however, they could not explain the stability of hedgehog lattices in MnGe. There the authors considered the so called 4th-order exchange and the chiral-chiral contributions allowing to properly describe MnGe \cite{grytsiuk2020topological}.

\subsubsection{Chiral bobbers}
When a skyrmion tube starts from a Bloch point and penetrates the remaining sample, the resulting magnetic object is called the chiral bobber~\cite{rybakov2015new,zheng2018experimental,ahmed2018chiral} (Fig.~\ref{fig:particles}j). It has been predicted~\cite{rybakov2015new} and observed experimentally by LTEM measurements~\cite{zheng2018experimental} in B20 materials. Fig.~\ref{fig:experiments}g shows a schematic representation of a pair of two chiral bobbers, as well as an experimental measurement of a Bloch point in Ni$_{80}$Fe$_{20}$.

One unique characteristics of a chiral bobber is that the location of the Bloch point, with respect to the sample's surface, is fixed~\cite{rybakov2015new}. It depends only on the ratio of exchange to DMI and is for example independent of the sample thickness~\cite{rybakov2015new}. Like a fishing bobber, the chiral bobber is `swimming' always close to the confinement of the sample.
For this reason, the chiral bobber must be seen as a distinct magnetic soliton despite its phenomenology as a discontinued skyrmion tube. Furthermore, this means that the volume which is occupied by a chiral bobber is fixed with respect to thickness variations, and so is the soliton's energy. For a skyrmion tube on the other hand, the occupied volume and the energy scale linearly with the thickness. For this reason, skyrmions in these materials become unstable above a critical sample thickness which is where chiral bobbers are energetically preferred~\cite{rybakov2015new}. It has been experimentally confirmed that chiral bobbers do not exist below a critical thickness~\cite{zheng2018experimental}.

In order to stabilize chiral bobbers, a specific experimental procedure had to be applied in Ref.~\cite{zheng2018experimental}: The magnetic field was \change{tilted} approximately $10^\circ$ out of the normal direction and then the sample was magnetized and demagnetized a few times. After the nucleation of the chiral bobbers and skyrmion tubes, the field was \change{tilted} back to the conventional out-of-plane direction. A possible nucleation mechanism is the formation of chiral bobbers from edge dislocations of the spin spiral phase, similar to the formation of skyrmion tubes. Skyrmion tubes and chiral bobbers have been distinguished by the LTEM phase shift which is proportional to the occupied volume of the magnetic object, i.\,e., chiral bobbers exhibit a weaker intensity than skyrmion tubes~\cite{zheng2018experimental}.

Even though a chiral bobber is not topologically protected (mathematically speaking the texture can be pushed out of the magnet via the confinement), it possesses a similar energy barrier compared to the skyrmion tube~\cite{rybakov2015new}. This allows for a coexistence of skyrmion tubes and chiral bobbers allowing to think of an alternative racetrack storage device where the '1' and '0' bits are encoded by the presence of a skyrmion tube or a chiral bobber, respectively (Fig.~\ref{fig:other}b). This solves the problem that the bits do not have to be located at precise positions. Instead, information can be encoded also as an irregular sequence~\cite{zheng2018experimental}. 

The two different objects can be read via their net magnetization or via their Hall signal. As has been shown numerically in Ref.~\cite{redies2019distinct}, chiral bobbers exhibit an increasing Hall conductivity upon increasing the sample thickness, while this signal remains constant for skyrmion tubes. 

The lack of topological protection may become problematic for the current-driven motion of chiral bobbers. As micromagnetic simulations of the STT scenario revealed, chiral bobbers do not only move along the sample but also the Bloch point propagates towards the edge along the tube direction until the chiral bobber has disappeared~\cite{kagawa2017current}. The SOT scenario remains to be investigated. 

\subsubsection{Hopfions}

A different way to transform a skyrmion tube into an innately three-dimensional object is to form a torus. These objects are known as hopfions (as in Fig.~\ref{fig:particles}l) and have been introduced back in 1931~\cite{hopf1964abbildungen}. They are topologically characterized by the Hopf invariant~\cite{whitehead1947expression,wilczek1983linking}
\begin{equation}
Q_H=-\change{\frac{1}{(4\pi)^2}}\int\vec{B}_\mathrm{em}(\vec{r})\cdot\vec{A}(\vec{r})\,\mathrm{d}^3r,
\end{equation}
where $\vec{B}_\mathrm{em}$ is the emergent field and $\vec{A}$ is the corresponding vector potential fulfilling the condition $\nabla\times\vec{A}=\vec{B}_\mathrm{em}$. 
In the simplest case, the cross-section texture is a bimeron with a topological charge of $\pm 1$ and the magnetization rotates once going around a circle of fixed radius around the hopfion's center (as in Fig.~\ref{fig:particles}l). This yields the Hopf number of $\pm 1$ as a product of the cross-section topological charge and the number of magnetization windings around the torus.

From a mathematical point of view, several types of hopfions can be constructed. For example, the cross section can have higher topological charges or the magnetization can rotate more than once going around a circle. Also, the Hopf number is increased by more complicated configurations, e.\,g., by linking multiple hopfions~\cite{rybakov2019magnetic}.

Hopfions have first been predicted in 1975 in a Skyrme-Faddeev model~\cite{korepin1975quantization}. Over the following years, hopfions have been found in hydrodynamics~\cite{kuznetsov1980topological}, electrodynamics~\cite{ranada1989topological} and other fields of physics, and recently they have been stabilized in micromagnetic simulations~\cite{sutcliffe2007vortex,tai2018static,liu2018binding,sutcliffe2018hopfions, rybakov2019magnetic}. The objects can be stabilized in chiral nanodisks due to the confinement~\cite{liu2018binding} and have even been stabilized without magnetic fields~\cite{rybakov2019magnetic}.

Their emergent electrodynamics are promising due to a globally compensated emergent field that is however finite locally (largest inside of the tube that forms the torus, as shown in Fig. \ref{fig:differentemergent}c). This toroidal emergent field leads to the deflection of current electrons perpendicular to the hopfion plane: One half of the object leads to a positive Hall resistance while the other half gives a negative signal. While both signals cancel on a global level, a local measurement of a hopfion in a potential racetrack device yields a unique signature as simulated in Ref.~\cite{gobel2020topological}. Furthermore, the topological Hall signal depends on the orientation of the hopfion bringing about the possibility to switch electrical signals (Fig.~\ref{fig:other}c): the Hall voltage always emerges perpendicularly to the hopfion plane. 
This brings about another interesting consequence. As for the bimeron, the emergent field $\vec{B}_\mathrm{em}$ of the hopfion (orange arrows in Fig. \ref{fig:differentemergent}c) is not parallel to the net magnetization $\vec{M}$ or to the stabilizing magnetic field $\vec{B}$ (both black arrows in Fig. \ref{fig:differentemergent}c). Therefore, the topological Hall effect is expected to enter a different tensor element of the resistivity compared to the ordinary and the anomalous Hall effects~\cite{gobel2020topological}. The topological Hall effect is expected to occur in a pure manner making it directly accessible and highly relevant for spintronic applications.

Likewise, when driven by torques, the hopfions move along the current direction and do not experience a skyrmion Hall effect~\cite{wang2019current}. The locally finite emergent field leads however to a tilting of the hopfion plane, as was first predicted in Ref.~\cite{gobel2020topological} and later confirmed by micromagnetic simulations~\cite{liu2020three}. On the one hand, if this effect is kept small, hopfions are promising for racetrack storage applications. On the other hand, the tilting mechanism may even be utilized for spintronic applications because it allows to switch between different hopfion configurations.

\section{Perspectives and conclusion} \label{sec:conclusion}

Summarizing this review, we have discussed the stability and the emergent electrodynamics of skyrmions and related alternative magnetic quasiparticles. We have classified the manifold of particles (Fig.~\ref{fig:overview}) in fundamental excitations (topologically trivial and non-trivial), variations of these excitations and extensions. Out of all fundamental excitations, skyrmions with an arbitrary helicity~\cite{okubo2012multiple,garlow2019quantification} and antiskyrmions~\cite{nayak2017magnetic} have been discussed as technologically relevant objects. The variations comprise topological excitations in a different magnetic background and the combination of multiple subparticles. Here we have discussed in detail bimerons~\cite{kharkov2017bound,gobel2018magnetic,gao2019creation}, biskyrmions~\cite{yu2014biskyrmion,gobel2019forming}, skyrmioniums~\cite{zhang2016control,zhang2018real}, antiferromagnetic skyrmions~\cite{barker2016static,legrand2019room},
and ferrimagnetic skyrmions~\cite{woo2017current,kim2017self}. 
`Extensions' means that all of these objects can be arranged in two dimensions (periodically and non-periodically) and can be continued along the third dimension. For the innately three-dimensional objects, we have laid focus on 
Bloch points~\cite{wild2017entropy,im2019dynamics},
chiral bobbers~\cite{rybakov2015new,zheng2018experimental} and hopfions~\cite{liu2018binding}. 

Out of the presented objects (Fig.~\ref{fig:particles}) antiferromagnetic skyrmion are often considered the optimal bits for spintronic applications. Their compensated magnetic texture allows to drive them by currents at enormously high velocities of up to several kilometers per second and the absence of the skyrmion Hall effect eradicates the problem of pinning of the bits at the edges. Furthermore, their local compensation of magnetization renders stray fields small allowing for a dense stacking of racetracks in a three-dimensional buildup.

Moreover, antiskyrmions seem to be highly promising. They are easy to detect and most experimental techniques working for the conventional skyrmions can be carried over. These objects can move parallel to an applied current allowing to consider antiskyrmions as the carriers of information in racetrack devices.
One further problem that occurs whenever the `0' and `1' bits of information are constituted by the presence or absence of a magnetic quasiparticle is that the positions of the bits cannot be precisely fixed. On the long time scale the information may become falsified for example due to the repulsive interaction between the objects. A solution may be delivered by using two distinct objects to constitute the bits because then irregular sequences are not problematic. A promising example is using skyrmions and antiskyrmions in systems with $D_{2d}$ symmetry~\cite{jena2020formation,jena2020elliptical,peng2020controlled,Gungordu2016, hoffmann2017antiskyrmions} (Fig.~\ref{fig:other}a). Also it is conceivable to use a three-dimensional approach with chiral bobbers and skyrmion tubes as bits of information~\cite{zheng2018experimental} (Fig.~\ref{fig:other}b).

Furthermore, it seems worthwhile to look into other applications of topologically non-trivial quasiparticles. 
Conventional skyrmions have been considered for utility in logic devices~\cite{zhang2015magnetic}, transistors~\cite{zhang2015magnetic2}, magnetic tunnel junctions~\cite{zhou2014reversible,kasai2019voltage,penthorn2019experimental}, nano-oscillators~\cite{Shen2019APL}, as microwave devices~\cite{wang2015driving} or magnonic devices~\cite{zhang2015all}, neuromorphic applications~\cite{huang2017magnetic,li2017magnetic}, as well as reservoir computing~\cite{prychynenko2018magnetic,pinna2018reservoir}, stochastic computing \cite{pinna2018skyrmion,zazvorka2019thermal} and quantum computing \cite{yang2016majorana,hals2016composite} etc. Future will tell if the alternative quasiparticles are favorable also in these regards. 

As a closing remark, we would like to mention that the presented characterization scheme is not \change{final}. \change{For example, very recently, it has been explored theoretically that skyrmions, antiskyrmions, skyrmioniums and other spin textures can carry so-called chiral kinks, i.\,e. defects of disfavored chirality within the spin texture \cite{kuchkin2020magnetic}. These kinks carry a topological charge and allow to realize new topological objects. Furthermore,} multiple skyrmions can in principle be combined to form new objects; e.\,g., a quadskyrmion instead of a biskyrmion. \change{Also}, one can combine multiple objects and change the magnetic background; antiferromagnetic versions of skyrmioniums~\cite{bhukta2018novel,obadero2019current}, skyrmion bags~\cite{rybakov2019chiral} and bimerons~\cite{fernandes2019skyrmions,shen2020,Li2020bimeron} have already been predicted. Since antiferromagnetic skyrmions have been observed just recently, the desire arises to find also antiferromagnetic versions of other nano-objects.\\
\\
\textbf{Acknowledgements}\\
This work is supported by SFB TRR 227 of Deutsche Forschungsgemeinschaft (DFG). O.A.T. acknowledges the support by the Australian Research Council (Grant No. DP20 0101027) and by the Cooperative Research Project Program at the Research Institute of Electrical Communication, Tohoku University.\\
\\
\textbf{Author contributions}\\
B.G. initiated the project and developed the scope and focus of the review with the help of I.M. and O.A.T.. The manuscript was written by B.G. with significant contributions from I.M. and O.A.T.. The figures were created by B.G..\\
\\

\bibliographystyle{model1-num-names}


\bibliography{short,MyLibrary}

\end{document}